\documentclass[11pt]{article}
\usepackage{jcappub}
\usepackage{bm}
\usepackage{color}
\usepackage{graphicx}

\newcommand{\be}{\begin{equation}}
\newcommand{\ee}{\end{equation}}
\newcommand{\een}{\end{subequations}}
\newcommand{\ben}{\begin{subequations}}
\newcommand{\beq}{\begin{eqalignno}}
\newcommand{\eeq}{\end{eqalignno}}

\newcommand{\lsim}{\mathrel{\mathop{\kern 0pt \rlap
      {\raise.2ex\hbox{$<$}}}\lower.9ex\hbox{\kern-.190em $ \sim$}}}
\newcommand{\gsim}{\mathrel{\mathop{\kern 0pt
      \rlap{\raise.2ex\hbox{$>$}}}\lower.9ex\hbox{\kern-.190em $\sim$}}}

\newcommand{\Ed}{E^{\prime}}

\title{Inelastic dark matter with spin--dependent couplings to protons
  and large modulation fractions in DAMA}

\author[a]{Stefano Scopel,}
\author[b]{Kook-Hyun Yoon}
\emailAdd{scopel@sogang.ac.kr}
\emailAdd{koreasds@naver.com}
\affiliation{Department of Physics, Sogang University, Seoul, South Korea}

\abstract{We discuss a scenario where the DAMA modulation effect is
  explained by a Weakly Interacting Massive Particle (WIMP) which
  upscatters inelastically to a heavier state and predominantly
  couples to the spin of protons. In this scenario constraints from
  xenon and germanium targets are evaded dynamically, due to the
  suppression of the WIMP coupling to neutrons, while those from
  fluorine targets are evaded kinematically, because the minimal WIMP
  incoming speed required to trigger upscatters off fluorine exceeds
  the maximal WIMP velocity in the Galaxy, or is very close to it. In
  this scenario WIMP scatterings off sodium are usually sensitive to
  the large--speed tail of the WIMP velocity distribution and
  modulated fractions of the signal close to unity arise in a natural
  way. On the other hand, a halo--independent analysis with more
  conservative assumptions about the WIMP velocity distribution allows
  to extend the viable parameter space to configurations where large
  modulated fractions are not strictly necessary. We discuss large
  modulated fractions in the Maxwellian case showing that they imply a
  departure from the usual cosine time dependence of the expected
  signal in DAMA. However we explicitly show that the DAMA data is not
  sensitive to this distortion, both in time and frequency space, even
  in the extreme case of a 100 \% modulated fraction. Moreover the
  same scenario provides an explanation of the maximum in the energy
  spectrum of the modulation amplitude detected by DAMA in terms of
  WIMPs whose minimal incoming speed matches the kinematic threshold
  for inelastic upscatters. For the elastic case the detection of such
  maximum suggests an inversion of the modulation phase below the
  present DAMA energy threshold, while this is not expected for
  inelastic scattering. This may allow to discriminate between the two
  scenarios in a future low--threshold analysis of the DAMA data.}

\begin{document}

\maketitle

\section{Introduction}
\label{sec:introduction}

The visible disk of our Galaxy is believed to be embedded in a halo of
Weakly Interacting Massive Particles (WIMPs). The DAMA
experiment\cite{dama} has been measuring for more than 15 years a
yearly modulation effect with a sodium iodide target consistent with
that expected due to the Earth rotation around the Sun from the
elastic scattering of WIMPs.  However, many experimental
collaborations using nuclear targets different from $NaI$ and various
background--subtraction techniques to look for WIMP--elastic
scattering (LUX\cite{lux}, XENON100\cite{xenon100},
XENON10\cite{xenon10}, KIMS\cite{kims,kims_modulation,kims2},
CDMS-$Ge$\cite{cdms_ge}, CDMSlite \cite{cdms_lite},
SuperCDMS\cite{super_cdms}, CDMS II\cite{cdms_2015},
SIMPLE\cite{simple}, COUPP\cite{coupp}, PICASSO\cite{picasso},
PICO-2L\cite{pico2l}, PICO-60\cite{pico60}) have failed to observe any
anomaly so far, implying severe constraints on the most popular WIMP
scenarios used to explain the DAMA excess, such as WIMP--nucleus
elastic scattering with a cross section proportional to the square of
the atomic mass number of the target or to the nuclear spin. In
particular the latter scenario consists in a WIMP fermionic particle
$\chi$ (either Dirac or Majorana) that recoils on the target nucleus
$T$ through its coupling to the spin $\vec{S}_N$ of nucleons
${\cal N}=(p,n)$:

\begin{equation} {\cal L}_{int}\propto \vec{S}_{\chi}\cdot
  \vec{S}_{\cal N}=c^p\vec{S}_{\chi}\cdot \vec{S}_p+c^n\vec{S}_{\chi}\cdot
  \vec{S}_n.
\label{eq:spin_dependent}
\end{equation}

One of the main motivations of the above interaction Lagrangian is the
fact that the most stringent bounds on the interpretation of the DAMA
effect in terms of WIMP--nuclei scatterings are obtained by detectors
using xenon (LUX\cite{lux}, XENON100\cite{xenon100}) and germanium
(CDMS\cite{cdms_ge,cdms_lite,super_cdms,cdms_2015}) whose spin is
mostly originated by an unpaired neutron, while both sodium and iodine
in DAMA have an unpaired proton: if the WIMP effective coupling to
neutrons $c^n$ is suppressed compared to that on protons $c^p$ this
class of bounds can be evaded
\cite{spin_n_suppression,spin_gelmini}. However this scenario is
constrained by droplet detectors (SIMPLE\cite{simple},
COUPP\cite{coupp}) and bubble chambers (PICASSO\cite{picasso},
PICO-2L\cite{pico2l},PICO-60\cite{pico60}) which all use nuclear
targets with an unpaired proton (in particular, they all contain
$^{19}F$, while SIMPLE contains also $^{35}Cl$ and $^{37}Cl$ and COUPP
and PICO-60 use also $^{127}I$).  As a consequence, this class of
experiments rules out an explanation of the DAMA effect in terms of
elastic WIMP--nucleus scatterings driven by the interaction
(\ref{eq:spin_dependent}) also for $c^n\ll c^p$ when standard
assumptions are made on the WIMP local density and velocity
distribution in our Galaxy\cite{spin_gelmini,pico2l}. This tension can
be relaxed if the spin--dependent scenario is generalized to a wider
class of interactions containing an explicit dependence of the cross
section on the exchanged momentum
\cite{spin_freitsis,spin_arina,eft_spin}.

In the present paper we wish to point out that an alternative approach
is possible to reconcile DAMA to fluorine detectors in the above
scenario with $c^n\ll c^p$: Inelastic Dark Matter
(IDM)\cite{inelastic}.  In this class of models a Dark Matter (DM)
particle $\chi$ of mass $m_{DM}$ interacts with atomic nuclei
exclusively by up--scattering to a second heavier state
$\chi^{\prime}$ with mass $m_{DM}^{\prime}=m_{DM}+\delta$. A peculiar
feature of IDM is that there is a minimal WIMP incoming speed in the
lab frame matching the kinematic threshold for inelastic upscatters
and given by:

\begin{equation}
v_{min}^{*}=\sqrt{\frac{2\delta}{\mu_{\chi N}}},
\label{eq:vstar}
\end{equation}

\noindent with $\mu{\chi N}$ the WIMP--nucleus reduced mass. This
quantity corresponds to the lower bound of the minimal velocity
$v_{min}$ (also defined in the lab frame) required to deposit a given
recoil energy $E_R$ in the detector:

\begin{equation}
v_{min}=\frac{1}{\sqrt{2 m_N E_R}}\left | \frac{m_NE_R}{\mu_{\chi N}}+\delta \right |.
\label{eq:vmin}
\end{equation}

\noindent The value of the recoil energy:

\begin{equation}
E_R=E_{R}^*=\delta\mu_{\chi N}/m_N,
\label{eq:estar}
\end{equation}
  
\noindent corresponding to $v_{min}$=$v_{min}^*$ usually coincides, as
in the case of the interaction (\ref{eq:spin_dependent}), to the
maximum of the signal.

The starting point of our analysis is the observation that, when the
WIMP mass is small enough and it is possible to assume that the DAMA
signal is only due to WIMP-sodium scatterings\footnote{In this case
  the KIMS experiment, containing $CsI$, is not sensitive to the DAMA
  effect due to its energy threshold\cite{kims,kims2}}, since $v_{min}^*$
decreases with the target mass $m_T$, it is larger for fluorine (with
mass $m_F\simeq$19.7 GeV) compared to sodium (with mass $m_{Na}\simeq$
21.4 GeV).  This difference in $v_{min}^*$ may seem to be small, due
to the mild dependence of $\mu_{\chi N}$ on $m_T$: however, precisely
when $m_{DM}$ is small the DAMA signal is produced by WIMPs in the
large-speed tail of their velocity distribution $f(\vec{v})$ in the
Galactic frame, where the signal can be highly sensitive to
$v_{min}$. This is what happens in the standard Isothermal Sphere
Model usually adopted to analyze direct detection data, i.e. a
Maxwellian representing a WIMP gas in thermal equilibrium with
r.m.s. velocity $v_{rms}\simeq \sqrt{3/2} v_{loc}\simeq$ 270 km/sec (with
$v_{loc}\simeq$ 220 km/sec the galactic rotation curve at the Earth's position\cite{v_loc})
and a velocity upper cut due to the escape velocity $v_{esc}$ (all quantities defined in the
Galactic reference frame).

In particular, indicating with $v_{min}^{*Na}$ and $v_{min}^{*F}$ the
values of $v_{min}^*$ for sodium and fluorine, the most extreme
situation is achieved when the WIMP mass $m_{DM}$ and the mass gap
$\delta$ imply the hierarchy:

\begin{equation}
v_{min}^{*Na}<v_{cut}^{lab}<v_{min}^{*F},
\label{eq:hierarchy}
\end{equation}

\noindent with $v_{cut}^{lab}$ the result of the boost in the lab rest
frame of some maximal value $v_{cut}$ beyond which the WIMP velocity
distribution in the Galactic rest frame vanishes (typically $v_{cut}$
is identified with the WIMP escape velocity $v_{esc}$): in this case
fluorine detectors turn outright blind to WIMP scatterings
\footnote{The COUPP and PICO-60 experiments contain also iodine and
  have an energy threshold significantly lower than KIMS: their case
  will be discussed separately in the next Section.}  while DAMA not
only remains sensitive to them, but it observes a modulation of the
signal which represents a very large fraction of the time--averaged
value, up to 100\%. In fact in Eq.(\ref{eq:hierarchy}) the boosted
value of the escape velocity $v_{esc}^{lab}$ oscillates back-and-forth
between the two constant quantities $v_{min}^{*Na}$ and $v_{min}^{*F}$
due to the annual change $\Delta v_{Earth}$ of the Earth velocity in
the Galactic rest frame between its maximal value in June and its
minimal value in December. When the WIMP mass is sufficiently small
$\Delta v_{Earth}\gsim$ $v_{min}^{*F}$-$v_{min}^{*Na}$ and the
amplitude of the oscillation is large enough to exceed the interval
between $v_{min}^{*F}$ and $v_{min}^{*Na}$: this means that in some
time interval centered in December $v_{min}^{*Na}>v_{esc}^{lab}$ and
also the signal in DAMA vanishes, implying a modulation fraction
approaching 100\%.  In addition to that $v_{min}^{*Na}$ belongs to the
$v_{min}$ interval explaining the DAMA signal and so the corresponding
energy $E_R^{*Na}$ belongs to the recoil energy interval where the
signal is measured; as it will be explained in Section
\ref{sec:modulation}, in the Maxwellian case this implies that for
$E_R$=$E_R^{*Na}$ the expected modulation amplitude has a maximum,
since the latter is a decreasing function of $v_{min}$ when $v_{min}$
is large (and in particular when it is close to the escape velocity,
as in our case); this feature might be in agreement with the energy
dependence of the modulation amplitudes measured by DAMA, which indeed
show a peak close to the threshold \cite{dama}. However this would not
imply a change in sign of the modulation amplitude at lower energies,
as predicted for elastic scattering of WIMPs of even lower
masses\cite{inverse_modulation}, allowing for a possible
discrimination of the two scenarios in future low--threshold analyses
of modulation data in DAMA \cite{dama_future}.

The qualitative outline summarized above will be confirmed by the
quantitative analysis contained in the following Sections, where we
will show that configurations $m_{DM}$--$\delta$ exist in the IDM
parameter space (close to the condition (\ref{eq:hierarchy}), but
allowing for some departure from it) where the DAMA annual modulation
effect can be explained by a WIMP signal in compliance with other
constraints. This will be achieved both by using a halo--independent
approach where the dependence of the expected signal on the specific
choice of $f(\vec{v})$ is factorized
\cite{factorization,mccabe_eta,gondolo_eta1,gondolo_eta2,gondolo_eta3,noi1},
and alternatively by fixing $f(\vec{v})$ to a Maxwellian distribution
and factorizing instead the WIMP--nucleon point--like cross section.

The very large modulation fractions implied by our scenario
necessarily imply a departure of the time--dependence of the signal
from the usual cosine functional form \cite{freese}. For this reason
we have also dedicated the last Section of our paper to a discussion
of how the ensuing distortions compare to the DAMA published data both
in time and in frequency space in the case of a Maxwellian
distribution, showing that the corresponding effects are below the
sensitivity of the experiment.

Our paper is organized as follows: in Section \ref{sec:fluorophobic}
we discuss the kinematic conditions that correspond to
Eq.(\ref{eq:hierarchy}) and combine them with the additional
requirements needed to evade the specific constraint of the COUPP
detector, which contains iodine targets and has a recoil energy
threshold lower than DAMA and KIMS; in Section \ref{sec:compatibility}
we describe the compatibility factors that we use in the subsequent
Sections to find the allowed configurations in our Scenario; Section
\ref{sec:analysis} contains our quantitative analysis; in Section
\ref{sec:modulation} we provide a discussion on how the large
modulation fractions arising in our scenario lead to a distortion of
the time dependence of the signal, and how this effect can be
accommodated by the DAMA data; Section \ref{sec:conclusions} contains
our conclusions. Information about our treatment of experimental data
is provided in Appendix \ref{app:exp}.

\section{The kinematics of fluorophobic IDM}
\label{sec:fluorophobic}

In the following we will assume a WIMP particle coupling to ordinary
matter via the spin--dependent interaction Lagrangian of
Eq.(\ref{eq:spin_dependent}) with $c^p\gg c^n$. For this reason in
this Section we will assume that bounds from detectors using xenon and
germanium targets are automatically evaded (although we will come back
to this issue in the numerical analysis of Section \ref{sec:analysis})
and will not consider them any further, concentrating instead on
detectors containing fluorine and iodine, which are sensitive to the
$c^p$ coupling.  As already pointed out, when the condition of
Eq.(\ref{eq:hierarchy}) is verified, droplet detectors and bubble
chambers turn blind to WIMP scatterings off fluorine targets since
WIMPs bound to the Galactic halo are not fast enough to trigger
upscatters of $\chi$ to the heavier state $\chi^{\prime}$, while at
the same time a population of WIMP particles with speeds below the
maximal velocity $v_{cut}$ still exists explaining the DAMA effect. 

Let's first consider the (usual) case when a specific choice for the
velocity distribution $f(\vec{v})$ is adopted (as in the case of a
Maxwellian). In this case $v_{cut}$ is equal to the escape velocity
$v_{esc}$, which can be determined from observation
\cite{vesc,vesc_salucci}. Using Eqs.(\ref{eq:vstar},\ref{eq:vmin}) one
gets the following conditions on the two free parameters $m_{DM}$ and
$\delta$:

\begin{eqnarray}
  \delta&>& \frac{1}{2} \left (v_{esc}^{lab}\right )^2 \mu_{\chi F} \label{eq:delta_bound_fluorine}\\
  \delta&<& \sqrt{2 m_{Na} E_{R,b}^{DAMA}}v^{lab}_{esc}-\frac{m_{Na}E_{R,b}^{DAMA}}{\mu_{\chi Na}}.
  \label{eq:delta_bound_dama}
\end{eqnarray}

\noindent Eq. (\ref{eq:delta_bound_fluorine}) corresponds to the
second inequality in Eq. (1.5), while Eq. (\ref{eq:delta_bound_dama})
corresponds to the requirement that the $v_{min}$ interval
corresponding to the DAMA signal falls below $v_{esc}$ (this latter
condition is more restrictive that the first inequality of
Eq.(\ref{eq:hierarchy})). In particular, in the equations above
$\mu_{\chi F}$, $\mu_{\chi Na}$ are the WIMP--fluorine and
WIMP--sodium reduced masses, respectively, while $E_{R,b}^{DAMA}$
corresponds to the experimental recoil energy boundary of the DAMA
signal that yields the stronger constraint on $\delta$. In this
Section we will neglect the effect of energy resolution and assume
that the DAMA modulated signal is concentrated in the range 2
keVee$\le E_{ee}^{DAMA} \le 4$ keVee for the equivalent--energy
$E_{ee}^{DAMA}$=$q E_R^{DAMA}$ (measured in keVee) and $q\simeq$ 0.3
the quenching factor on sodium. This implies explicitly
$E_{R,b}^{DAMA}\simeq$ 6.7 keV in Eq.(\ref{eq:delta_bound_dama}).

\begin{figure}
\begin{center}
\includegraphics[width=0.49\columnwidth, bb=73 193 513 636]{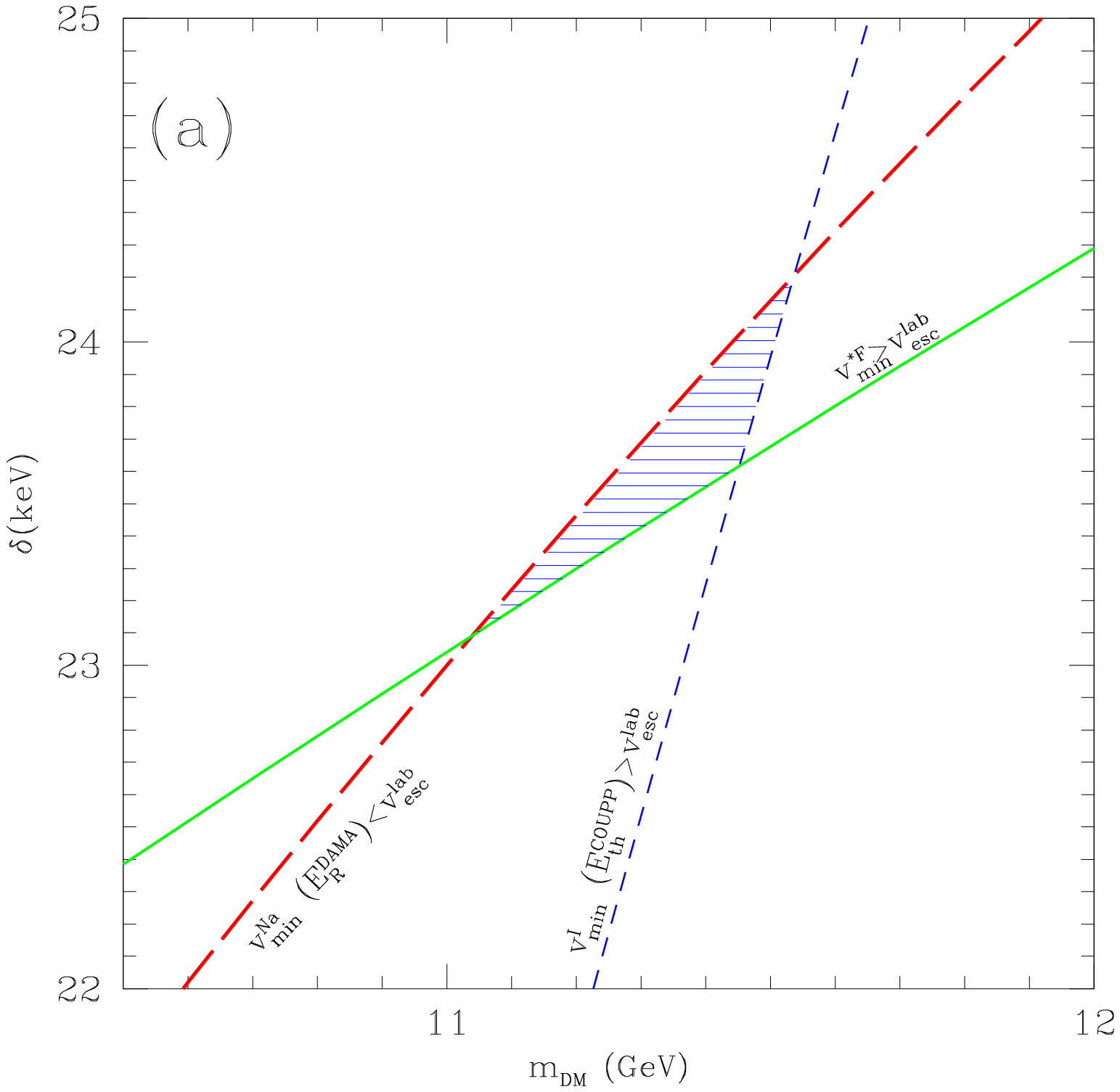}
\includegraphics[width=0.49\columnwidth, bb=73 193 513 636]{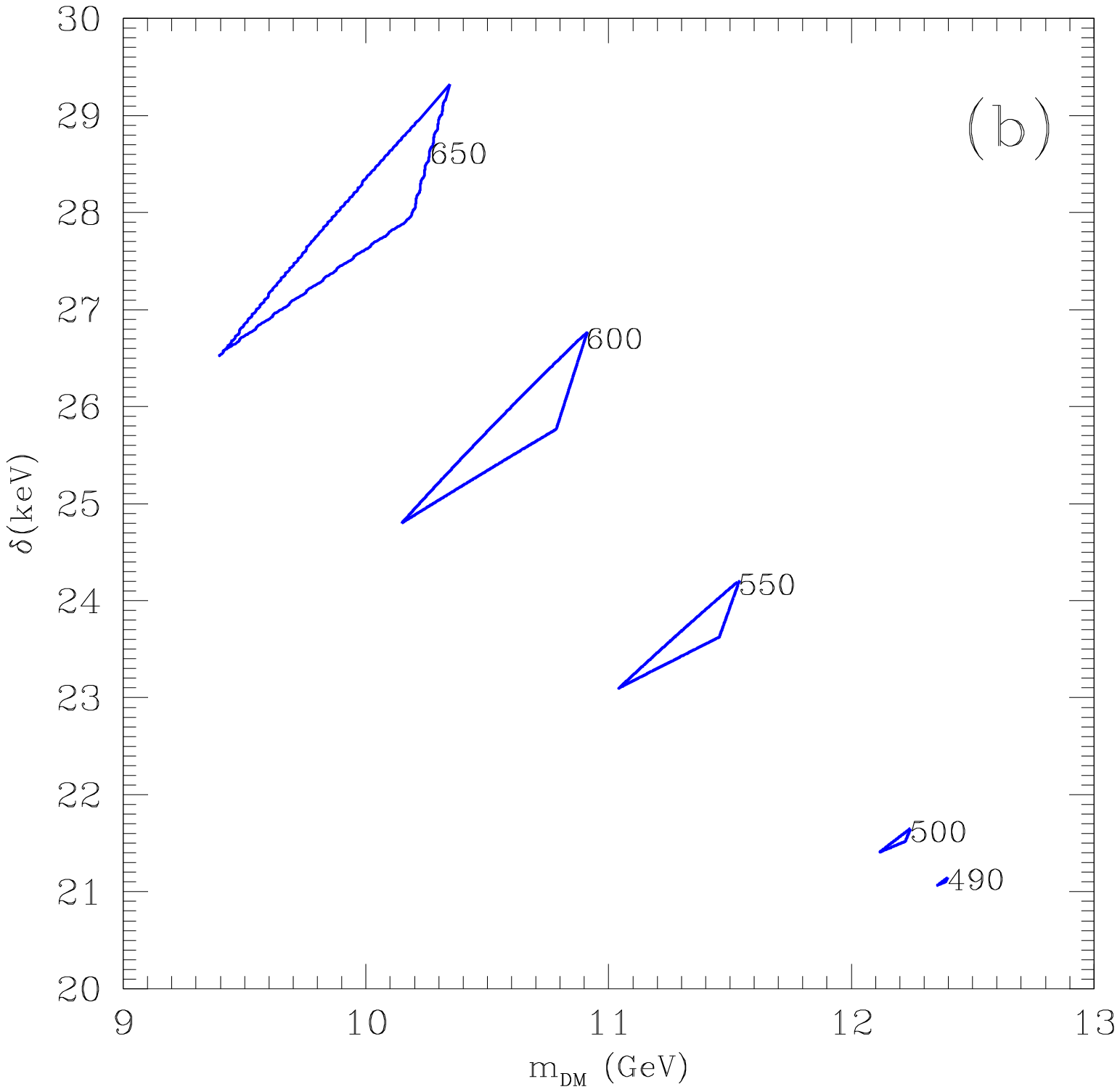}
\end{center}
\caption{{\bf (a)} The horizontally (blue) hatched area represents
  values of the IDM parameters $m_{DM}$ and $\delta$ for which the
  conditions of
  Eqs. (\ref{eq:delta_bound_fluorine},\ref{eq:delta_bound_dama},\ref{eq:delta_bound_coupp})
  are verified when $v_{esc}$=550 km/sec. {\bf (b)} Different
  determinations of the same region are shown for the indicated values
  of $v_{esc}$. }
\label{fig:mchi_delta_na_f_i}
\end{figure}

\begin{figure}
\begin{center}
\includegraphics[width=0.49\columnwidth, bb=73 193 513 636]{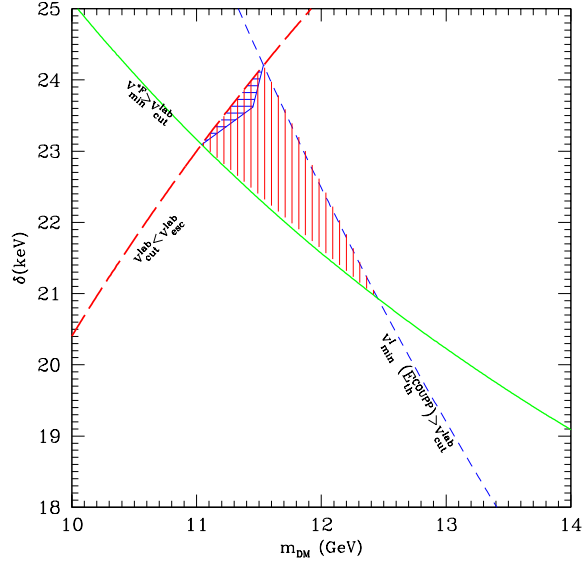}
\end{center}
\caption{The vertically (red) hatched region represents the IDM
  parameter space where the conditions of
  Eqs. (\ref{eq:delta_bound_fluorine},\ref{eq:delta_bound_dama},\ref{eq:delta_bound_coupp})
  are verified with $v_{esc}^{lab} \rightarrow v_{cut,DAMA}^{lab}$
  equal to the highest value of $v_{min}$ for which the DAMA effect is
  present (see text).  The additional condition $v_{cut,DAMA}^{lab}\le
  v_{esc}^{lab}$ is also enforced. The horizontally (blue) hatched
  area is the same shown in Fig.\ref{fig:mchi_delta_na_f_i}(a).}
\label{fig:mchi_delta_v_cut}
\end{figure}

The two curves corresponding to Eqs.(\ref{eq:delta_bound_fluorine})
and (\ref{eq:delta_bound_dama}) are shown with the the solid (green)
line and the thick (red) dashes, respectively in
Fig.\ref{fig:mchi_delta_na_f_i}(a) for $v_{esc}$=550 km/sec . In
particular, by assuming the value $v_{Sun}$=232 km/sec for the velocity
of the Solar system with respect to the WIMP halo this corresponds to
$v_{esc}^{lab}$=782 km/sec in the lab rest frame. In this case the
corresponding region of the $m_{DM}$--$\delta$ plane is between the
two curves for $m_{DM}\gsim$ 11 GeV.

For the low values of $m_{DM}$ shown in
Fig.\ref{fig:mchi_delta_na_f_i}(a) it is straightforward to show that
in order for scatters off iodine both in DAMA and in KIMS to be above
the corresponding thresholds large values of $v_{min}\gsim 900\,
\mbox{km/sec}> v_{esc}^{lab}$ are required. However, among bubble
chamber detectors both COUPP \cite{coupp} and PICO--60 \cite{pico60}
use trifluoroiodomethane targets (CF$_3$I) which also contain
iodine. Both experiments, which have not observed any excess in their
data, have energy thresholds substantially lower than DAMA and KIMS:
for COUPP $E_{th}^{COUPP}$=7.8 keV, while for PICO--60
$E_{th}^{PICO-60}$=7 keV, yielding some constraints to our
scenario. Since only a very small fraction of the livetime of the
PICO-60 was operated with a threshold lower than COUPP (about 1\% of
the total \cite{pico60}) the ensuing combined constraint is driven by
COUPP, as will be shown in our numerical analysis. The COUPP bound is
evaded when:

\begin{eqnarray}
  &&v_{min}(E^{COUPP})>v_{esc}^{lab}  \nonumber \\
  &&\Rightarrow \delta>\sqrt{2 m_I E^{COUPP}_{th}} v_{esc}^{lab}-\frac{m_I E^{COUPP}_{th}}{\mu_{\chi I}},\label{eq:hierarchy_coupp}
\label{eq:delta_bound_coupp}
\end{eqnarray}

\noindent with $m_I$ the iodine nucleus mass and $\mu_{\chi I}$ the
WIMP--iodine reduced mass. The corresponding curve for $v_{esc}$=550
km/sec is shown in Fig.\ref{fig:mchi_delta_na_f_i}(a) with the (blue)
dashed line.

The overlapping of the three regions corresponding to the conditions
(\ref{eq:delta_bound_fluorine},\ref{eq:delta_bound_dama},\ref{eq:delta_bound_coupp})
in the $m_{DM}$-$\delta$ plane yields the horizontally (blue) hatched
area in Fig.\ref{fig:mchi_delta_na_f_i}(a). In Figure
\ref{fig:mchi_delta_na_f_i}(b) we have repeated the same procedure
with different values of $v_{esc}$, showing that such bound contours
exist in a range of $v_{esc}$ encompassing those commonly used in the
literature\cite{vesc,vesc_salucci} \footnote{However, in
  self--truncated Isothermal models \protect\cite{self_truncated} that
  take into account the modifications of the function $f(\vec{v})$ due
  to the finite size of our Galaxy the r\"ole of the escape velocity
  in the calculation of the WIMP direct detection rate is played by a
  maximal speed $v_{max}\lsim$ 430 km/sec well below the range of
  $v_{esc}$ shown in Figure \protect\ref{fig:mchi_delta_na_f_i}(b).}.

An alternative and more conservative approach to assess the
compatibility of the DAMA result with the constraints from other
experiments consists in assuming instead
$v_{cut}^{lab}$=$v_{cut,DAMA}^{lab}$ with $v_{cut,DAMA}^{lab}$ the
highest value of $v_{min}$ for which the DAMA effect is
present\cite{eft_spin}. As discussed in Section
\ref{sec:compatibility}, in this way the halo function defined in
Eq.(\ref{eq:eta}) is the one yielding the smaller possible count rates
in other detectors in compliance with its minimal required conditions
(\ref{eq:eta_conditions}). This implies the substitution
$v_{esc}^{lab}\rightarrow v^{lab}_{cut,DAMA}(m_{DM},\delta)$ in
Eqs.(\ref{eq:delta_bound_fluorine},\ref{eq:delta_bound_dama},\ref{eq:delta_bound_coupp})
with the additional constraint $v_{cut,DAMA}^{lab}\le v_{esc}^{lab}$:
the ensuing range in the IDM parameters is shown in
Fig.\ref{fig:mchi_delta_v_cut} by the vertically (red) hatched region.
Again, here and in the following we adopt for $v_{esc}$ the reference
value $v_{esc}$=550 km/sec.

The conditions dictated by
(\ref{eq:delta_bound_fluorine},\ref{eq:delta_bound_dama},\ref{eq:delta_bound_coupp})
involve only kinematics, and are not strictly necessary. Indeed, as it
will be shown in the quantitative analysis of Section
\ref{sec:analysis}, both in the case of a Maxwellian velocity
distribution and for a halo--independent approach the allowed regions
in the IDM parameter space extend outside the bound regions shown in
Figs.\ref{fig:mchi_delta_na_f_i}(a) and \ref{fig:mchi_delta_v_cut}
(particularly so in the latter case) when also dynamical and
experimental considerations are taken into account. However, the
kinematic ranges discussed in this Section can be considered as the
starting seeds for a more comprehensive search of allowed IDM models
in our scenario.

\section{Compatibility factors}
\label{sec:compatibility}

In this Section we introduce the procedures that will be used in a
quantitative way in Section \ref{sec:analysis} to discuss the
compatibility with other constraints of an interpretation of the DAMA
effect in terms of our scenario.

The halo--independent approach \cite{factorization} consists in
writing the expected event rate for the WIMP--nucleus scattering
process in the observed detected energy interval $E_1^{\prime}\le
E^{\prime}\le E_2^{\prime}$ as \cite{mccabe_eta,gondolo_eta1,gondolo_eta2,gondolo_eta3, noi1}:

\begin{equation} 
R_{[\Ed_1, \Ed_2]}=\int_{0}^{\infty} d\,v_{min}
\tilde{\eta}(v_{min}) {\cal R}_{[\Ed_1,\Ed_2]}(v_{min}).
\label{eq:r_compact_eta}
\end{equation}

\noindent In the equation above the detected energy $E^{\prime}$
represents the fraction of the true nuclear recoil energy $E_R$ which
is actually measured in a given experiment (taking into account a
possible quenching factor $<$ 1 for ionization and scintillation)
after convolution with the experimental energy resolution,

\begin{equation}
\tilde{\eta}(v_{min})=\frac{\rho_{DM}\sigma_{ref}}{m_{DM}}\int_{v_{min}} d^3 \vec{v}_T \frac{f(\vec{v}_T)}{v_T}, 
\label{eq:eta}
\end{equation}

\noindent is the {\it halo-function} containing all the dependence of
the expected rate on the astrophysical assumptions for the WIMP halo
($\rho_{DM}$ represents the WIMP mass density in the neighborhood of
the Sun while $f(\vec{v_T})$ is the WIMP velocity distribution in the
target rest frame) while the response function $\cal R$ contains the
spin--dependent nuclear form factor and is specific for every
detector, since it includes the effects due to the energy resolution
and experimental acceptances. In the following we will adopt the form
factors as calculated in \cite{haxton1,haxton2} (although at the
beginning of Section \ref{sec:analysis} we will comment on different
determinations in the case of xenon).  For the other details regarding
the response function $\cal R$ we refer to Ref. \cite{eft_spin}: in
particular, the spin--dependent case of the interaction Lagrangian
(\ref{eq:spin_dependent}) corresponds to setting $c_k^{\tau}$=0 for
$k\ne$ 4 and $c^{0,1}_4$=($c^p\pm c^n$)/2 in Eqs. (4.5, A1) of
Ref. \cite{eft_spin}.

In Eq.(\ref{eq:eta}) $\sigma_{ref}$ represents a reference cross
section, defined in the limit of vanishing transferred momentum, that
we take as $\sigma_{ref}=(c^p)^2 \mu^2_{\chi N}/\pi$ \footnote{The
  WIMP --proton cross section corresponds to 3/16 $\sigma_{ref}$.}. In
this way the response function ${\cal R}$ will depend on the ratio
$r\equiv c^n/c^p$\footnote{In the following we will restrict our
  analysis to real values of $c^n$ and $c^p$, although in the case of
  inelastic scattering they can be complex.}.

Following the same procedure, it is also possible to introduce a
modulated halo function $\tilde{\eta}_1$ defined as:

\begin{equation}
\tilde{\eta}_1(v_{min})\equiv\left [\tilde{\eta}(v_{min},t=t_{max})-\tilde{\eta}(v_{min},t=t_{min})\right ]/2,
\label{eq:eta_modulation}
\end{equation}

\noindent with $t_{max}$ and $t_{min}$ the times of the year
corresponding to the maximum and to the minimum of the Earth's
velocity in the Galactic rest frame. In this way, the modulated
amplitudes measured by DAMA can be expressed, in analogy to
Eq.(\ref{eq:r_compact_eta}), as:

\begin{equation} 
\Delta R_{[\Ed_1, \Ed_2]}=\int_{0}^{\infty} d\,v_{min}
\tilde{\eta}_1(v_{min}) {\cal R}_{[\Ed_1,\Ed_2]}(v_{min}).
\label{eq:r_compact_eta1}
\end{equation}

The two halo functions $\tilde{\eta}$ and $\tilde{\eta}_1$ are subject
to the very general conditions:

\begin{eqnarray}
&\tilde{\eta}(v_{min,2})\le\tilde{\eta}(v_{min,1})  & \mbox{if $v_{min,2}> v_{min,1}$},\nonumber\\
& \tilde{\eta}_1\le\tilde{\eta}  & \mbox{at the same $v_{min}$},\nonumber\\
& \tilde{\eta}(v_{min} \ge v_{esc}^{lab})=0. &
\label{eq:eta_conditions}
\end{eqnarray}

The first condition descends from the definition (\ref{eq:eta}), that
implies that $\tilde{\eta}(v_{min})$ is a decreasing function of
$v_{min}$. The second is a consequence of the fact that
$\tilde{\eta}_1$ is the modulated part of $\tilde{\eta}$. The last
condition reflects the requirement that the WIMPs are gravitationally
bound to our Galaxy. As already done in Section \ref{sec:fluorophobic}
in the following we will assume that the WIMP halo is at rest in the
Galactic rest frame and we will adopt as the escape velocity of WIMPs
in the lab rest frame $v_{esc}^{lab}$=$v_{esc}^{Galaxy}$+$v_{Sun}$,
where $v_{esc}^{Galaxy}$=550 km/sec and $v_{Sun}$=232 km/sec the velocity
of the Solar system with respect to the WIMP halo.

The halo--independent method exploits the fact that $R_{[\Ed_1,
  \Ed_2]}$ and $\Delta R_{[\Ed_1, \Ed_2]}$ depend on $f(\vec{v_T})$
only through the minimal speed $v_{min}$ (given in
Eq.(\ref{eq:vmin})) that the WIMP must have to deposit at least
$E_R$. By mapping recoil energies $E_R$ (and so detected energies
$E^{\prime}$) into same ranges of $v_{min}$ the dependence on
$\eta(v_{min})$ and so on $f(\vec{v_T})$ cancels out in the ratio of
expected rates on different targets. In this way experimental
count--rates can be exploited to get direct information on the unknown
halo functions $\tilde{\eta}$ and $\tilde{\eta}_1$. In particular,
given an experiment with detected count rate $N_{exp}$ in the energy
interval $E_1^{\prime}<E^{\prime}<E_2^{\prime}$ the
combination\cite{gondolo_eta1}:

\begin{equation}
  <\tilde{\eta}>=\frac{\int_{v_{min}^*}^{\infty} d v_{min} \tilde{\eta}(v_{min}) {\cal R}_{[E_1^{\prime},E_2^{\prime}]}(v_{min})}
  {\int_{v_{min}^*}^{\infty} d v_{min} {\cal R}_{[E_1^{\prime},E_2^{\prime}]}(v_{min})}=\frac{N_{exp}}{\int_{v_{min}^*}^{\infty} d v_{min} {\cal R}_{[E_1^{\prime},E_2^{\prime}]}(v_{min})},
\label{eq:eta_bar_vt}
\end{equation}

\noindent can be interpreted as an average of the function
$\tilde{\eta}(v_{min})$ in an interval $v_{min,1}<v_{min}<v_{min,2}$
(an analogous argument holds for $\tilde{\eta}_1(v_{min})$, when the
detected count rate represents a modulation amplitude as those
measured by DAMA). The $v_{min}$ interval is defined as the one where
the response function ${\cal R}$ is ``sizeably'' different from zero
(we will conventionally take the interval
$v_{min}[E_R(E_{ee,1})]<v_{min}<v_{min}[E_R(E_{ee,2})]$ with
$E_{ee,1}=E^{\prime}_1-\sigma_{rms}(E^{\prime}_1)$,
$E_{ee,2}=E^{\prime}_2+\sigma_{rms}(E^{\prime}_2)$ and
$\sigma_{rms}(E^{\prime})$ the energy resolution).

Following the procedure outlined above it is straightforward, for a
given choice of the DM parameters, to obtain estimations
$\bar{\tilde{\eta}}_{1,i}^{DAMA-Na}$ of the modulated halo function
$\tilde{\eta}_1(v_{min})$ averaged in appropriately chosen $v_{min,i}$
intervals mapped from the DAMA experimental annual modulation
amplitudes. Using the condition
$\tilde{\eta}_1(v_{min})\le\tilde{\eta}(v_{min})$ this allows to get
lower bounds on the $\tilde{\eta}(v_{min})$ function, which can be
compared to the upper bounds $\bar{\tilde{\eta}}_{j,lim}$ on the same
quantity derived from the data of the experiments that have reported
null results\footnote{In the case of inelastic scattering the
  correspondence between $v_{min}$ and $E_R$ is no longer
  one--to--one, so that some value of $v_{min}$ may correspond to two
  values of $E_R$: we avoid this by binning the energy intervals in
  such a way that, for each experiment, $E_R^*$ as defined in
  Eq.(\ref{eq:estar}) corresponds to one of the bin
  boundaries\cite{noi1}.}.

Quantitatively, for a given choice of the WIMP mass $m_{DM}$, of the
mass difference $\delta$ and of the ratio $r=c^n/c^p$, the
compatibility between DAMA and all the other results can be assessed
introducing the following compatibility ratio \cite{noi2}:

\begin{equation}
  {\cal D}(m_{DM},r,\delta) \equiv \max_{i\in \mbox{signal}}
\left (\frac{\bar{\tilde{\eta}}_i^{DAMA-Na}+\sigma_i}{\min_{j\le i}\bar{\tilde{\eta}}_{j,lim}} \right ),
\label{eq:compatibility_factor_eta_i}
\end{equation}

\noindent where $\sigma_i$ represents the standard deviation on
$\bar{\tilde{\eta}}_i^{DAMA-Na}$ as estimated from the data, $i\in
\mbox{signal}$ means that the maximum of the ratio in parenthesis is
for $v_{min,i}$ corresponding to the DAMA excess, while, due to the
fact that the function $\tilde{\eta}$ is non--decreasing in all
velocity bins $v_{min,i}$, the denominator contains the most
constraining bound on $\tilde{\eta}$ for $v_{min,j}\le v_{min,i}$.
The latter minimum includes all available bounds from scintillators,
ionizators and calorimeters (see Appendix \ref{app:exp} of the present
paper and Appendix B of Ref.\cite{eft_spin} for a summary of the
experimental inputs used in our analysis). Specifically,
compatibility between DAMA and the constraints included in the
calculation of Eq.(\ref{eq:compatibility_factor_eta_i}) is ensured if
${\cal D}<1$.  Note that requiring the
compatibility factor of Eq.(\ref{eq:compatibility_factor_eta_i}) (with
the "+" sign in the numerator) to be below unity corresponds to
accepting only configurations where upper bounds do not cut into the
DAMA signal region altogether.  Our choice has the advantage to
automatize the search of compatible regions in a simple and easily
reproducible way and to be a conservative one without resorting to the
more involved statistical combination of inhomogeneous data sometimes
affected by large systematic uncertainties. The ensuing results should
be considered as indicative and depend on the assumed C.L.

The above procedure cannot be applied to bubble chambers and droplet
detectors, which are only sensitive to the energy threshold, because
in this case it is not possible to map the corresponding bounds to
arbitrary velocity bins. Moreover, they all contain different nuclear
targets ($C_2 Cl F_5$ for SIMPLE\cite{simple}, $C F_3 I$ for
COUPP\cite{coupp} and PICO-60\cite{pico60}, $C_4 F_{10}$ for
PICASSO\cite{picasso} and $C_3 F_8$ for PICO-2L\cite{pico2l} so that
it is in general not possible to factorize the $\tilde{\eta}$
function in a specific range of $v_{min}$.  In Ref.\cite{eft_spin} we introduced an alternative
procedure to handle this class of experiments, which we summarize
here: i) we use the experimental DAMA modulation--amplitudes to get a
conservative piecewise estimation $\tilde{\eta}^{est}_1(v_{min})$ of
the minimal $\tilde{\eta}_1$ modulated halo function compatible to the
signal (see for instance the (blue) dots --short dashes in
Figs.\ref{fig:vmin_eta_benchmark_1} and
\ref{fig:vmin_eta_benchmark_2_3}); ii) we obtain the corresponding
estimation of the unmodulated part $\tilde{\eta}^{est}(v_{min})$ by
requiring that it is a decreasing function of $v_{min}$ with
$\tilde{\eta}^{est}(v_{min})\ge \tilde{\eta}^{est}_1(v_{min})$ (an
explicit example is provided by the (red) dots --long dashes in
Figs.\ref{fig:vmin_eta_benchmark_1} and
\ref{fig:vmin_eta_benchmark_2_3}); iii) in compliance with
(\ref{eq:eta_conditions}) and with the goal of obtaining a
conservative bound, we require that the function $\tilde{\eta}$ is the
{\it minimal} one able to explain the DAMA effect, so we assume (as
discussed in the second part of Section \ref{sec:fluorophobic}) that
$\tilde{\eta}^{est}(v_{min}>v_{cut,DAMA}^{lab})$=0, with
$v_{cut,DAMA}^{lab}$ the highest value of $v_{min}$ for which the DAMA
effect is present; iv) we then use $\tilde{\eta}^{est}(v_{min})$ to
directly calculate for each experiment among $k$=SIMPLE, COUPP,
PICO-60, PICASSO and PICO-2L and for each energy threshold $E_{th,i}$
the expected number of WIMP events $N_{k,i}^{expected}$ and compare it
to the corresponding 90\% C.L. upper bound $N_{k,i}^{bound}$ (see
Appendix B of Ref\cite{eft_spin} and Appendix \ref{app:exp} of the
present paper for further details).

As pointed out in the Introduction, in our scenario it is quite
natural that $E_R^{*Na}$ falls in the energy interval where the DAMA
modulation amplitude has been detected. This implies the existence of
pairs of energy bins in DAMA that are mapped into the same range for
$v_{min}$: in this case one gets two different determinations
$\bar{\tilde{\eta}}_{1,1}$ and $\bar{\tilde{\eta}}_{1,2}$ of the
modulated halo function in that particular range of $v_{min}$ which
must be mutually compatible; in order to ensure this we apply the
following {\it shape test}\cite{halo_independent_inelastic}:

\begin{equation}
\Delta_{ST}\equiv \frac{\left |\bar{\tilde{\eta}}_{1,1}-\bar{\tilde{\eta}}_{1,2}\right |}{\sqrt{\sigma_1^2+\sigma_2^2}}\le 1.96,
\label{eq:shape_test}
\end{equation}
\noindent at the 95 \% C.L.

Although in the halo-independent analysis the halo function is
directly fixed to the observed values of the modulation amplitudes,
there exist two cases when the ensuing total signal is not guaranteed
to correspond to what observed by DAMA.  The first case is when part
of the DAMA region falls beyond $v^{lab}_{cut}$; the second case
corresponds to the situation when two DAMA energy bins are mapped into
the same velocity range and we calculate the corresponding value of
the halo function as the combination of the two determinations. In
both cases, to ensure that a particular choice of the IDM parameters
provides an acceptable explanation of the DAMA effect we require that
the total modulated rate $\Delta R_{2,4}$ in the energy interval 2
keVee$\le E^{\prime}\le$ 4 keVee obtained using such halo-function
determinations exceeds the 95\% C.L. lower bound $\Delta
R_{2,4,min}^{DAMA}$=0.028 events/kg/day on the corresponding observed
quantity\cite{dama}.

Then, a straightforward generalization of the compatibility
factor of Eq.(\ref{eq:compatibility_factor_eta_i}) is:

\be {\cal D}(m_{DM},r,\delta)\rightarrow \max \left ({\cal
  D}(m_{DM},r,\delta), \frac{N_{k,i}^{expected}}{N_{k,i}^{bound}},\frac{\Delta_{ST}}{1.96}, \frac{\Delta R_{2,4}}{\Delta R_{2,4,min}^{DAMA}} \right ).
\label{eq:compatibility_factor_generilized}
\ee

In Section \ref{sec:analysis} we will also test the compatibility of
the DAMA modulation effect when the WIMP velocity distribution is
fixed to a standard Maxwellian. This is the usual approach adopted in
the literature, where in the expression of the expected rate knowledge
of $f(\vec{v})$ allows to directly factorize the cross section
$\sigma_{ref}$ . In this case for a fixed value of the WIMP mass (and
for a choice of the local WIMP density $\rho_{DM}$) the whole
experimental spectrum allows to get a single upper bound, or in case
of an excess, a single estimation of $\sigma_{ref}$. As far as the
DAMA modulation effect is concerned, we will estimate an interval
$\sigma_{DAMA}^{min}\le\sigma_{eff}\le\sigma_{DAMA}^{max}$ for the
cross section by minimizing a $\chi$--square of the $\tilde{\eta}_1$
function through the corresponding experimental estimations (in this
case $\sigma_{eff}$ is just a normalization factor of
$\tilde{\eta}_1$). In the case of null experiments we will use the
same energy bins used for the halo--independent approach, and add one
last bin containing the whole experimental range analyzed in the
experiment. For each experiment $k$ and each energy bin $i$ we will
then calculate the corresponding upper bound $\sigma_{ki}^{bound}$.

Also in the Maxwellian analysis we apply a quality
check on the corresponding prediction for the modulation amplitudes.
In particular, when $\sigma_{ref}$ is fixed to its best--fit value  
we require that the $p$--value of the minimal
$\chi$--square exceeds $p_{min}$=0.05.

Then, in the following Section we will adopt for the Maxwellian case
the compatibility factor:

\begin{equation} 
{\cal D}_{Maxwellian}(m_{DM},r,\delta) 
\equiv \max\left [ \max_{k,i}
\left (\frac{\sigma_{DAMA}^{min}}{\sigma_{ki}^{bound}} \right ),\frac{p_{min}}{p} \right ].
\label{eq:compatibility_factor_maxwellian}
\end{equation}

\section{Analysis}
\label{sec:analysis}

In this Section we wish to discuss the compatibility with other
constraints of an interpretation of the DAMA effect in terms of the
IDM scenario introduced in Section \ref{sec:introduction}, with
couplings to nuclei driven by Eq.(\ref{eq:spin_dependent}) and $c^n\ll
c^p$. In order to do that, we will explore the IDM parameter space
($m_{DM}$,$\delta$,$r\equiv c^n/c^p$) calculating both the
compatibility factor ${\cal D}$ defined in
Eq.(\ref{eq:compatibility_factor_generilized}) (with no assumptions on
the halo function $\tilde{\eta}$ besides the minimal ones listed in
Eq(\ref{eq:eta_conditions})), and ${\cal D}_{Maxwellian}$ as defined
in Eq.(\ref{eq:compatibility_factor_maxwellian}), where, instead, a
Maxwellian distribution is assumed for $f(\vec{v})$.

\begin{figure}
\begin{center}
\includegraphics[width=0.49\columnwidth, bb=73 193 513 636]{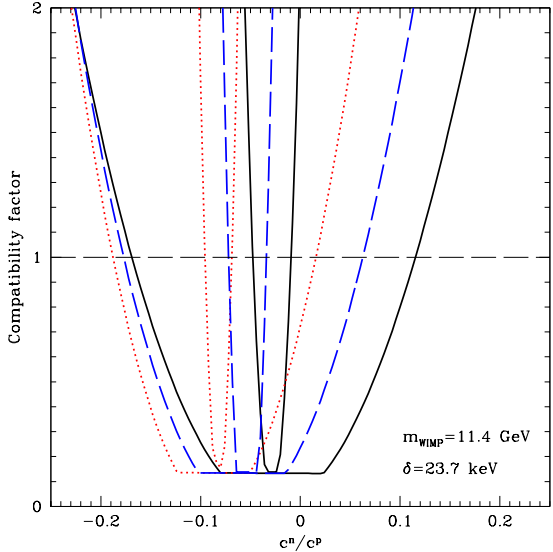}
\end{center}
\caption{The compatibility factors defined in
  Eqs.(\ref{eq:compatibility_factor_generilized}) and
  (\ref{eq:compatibility_factor_maxwellian}) are plotted as a function
  of the ratio $c^n/c^p$ for $m_{DM}$=11.4 GeV, $\delta$=23.7 keV
  (benchmark point $P_1$ in
  Fig.\ref{fig:mchi_delta_full_analysis}). Solid lines: ${\cal D}$ and
  ${\cal D}_{Maxwellian}$ where the xenon spin--dependent nuclear form
  factor is evaluated using Ref \cite{haxton1,haxton2}; (red) dots:
  the same curves for the Bonn-A xenon spin--dependent form
  factor\cite{spin_form_factors}; (blue) dashes: Nijmegen xenon
  spin--dependent form factor\cite{spin_form_factors}. In all cases
  the shallower curve represents ${\cal D}$ while the steeper one
  shows ${\cal D}_{Maxwellian}$.}
\label{fig:r_d}
\end{figure}

\begin{figure}
\begin{center}
\includegraphics[width=0.49\columnwidth, bb=73 193 513 636]{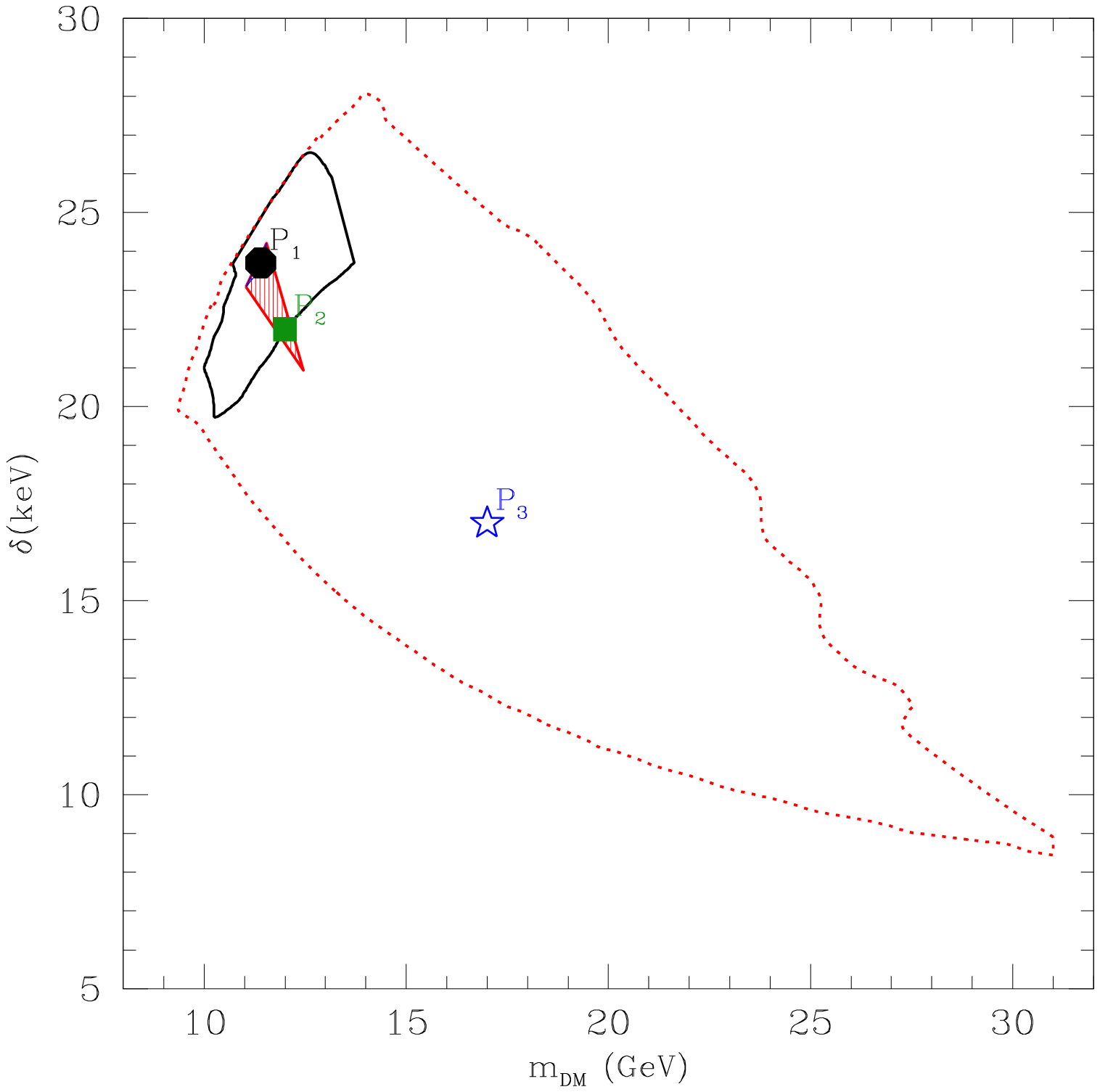}
\includegraphics[width=0.49\columnwidth, bb=73 193 513 636]{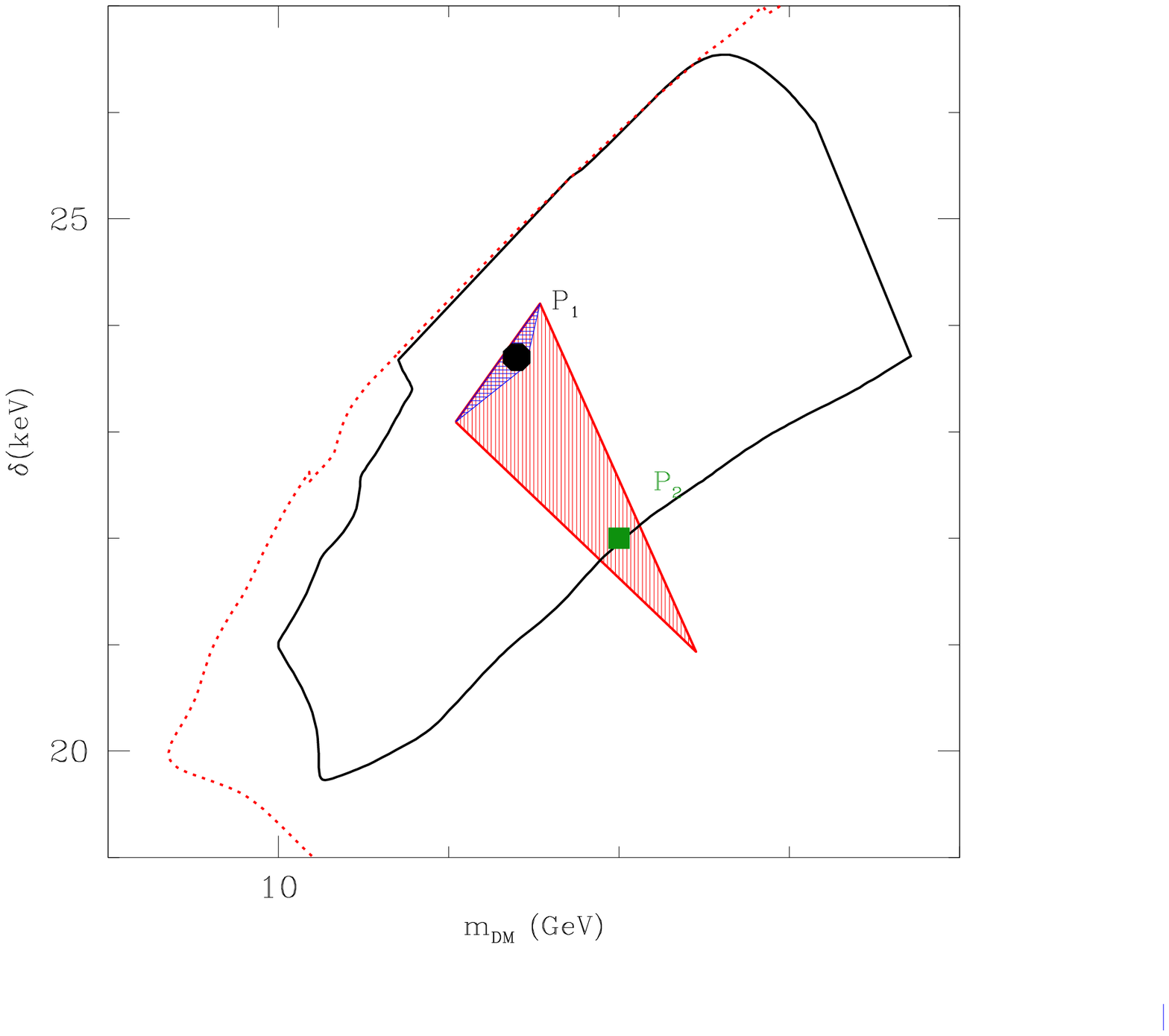}
\end{center}
\caption{Contour plots of the compatibility factors ${\cal D}$
  (\ref{eq:compatibility_factor_generilized}) and ${\cal
    D}_{Maxwellian}$ (\ref{eq:compatibility_factor_maxwellian}) in the
  plane $m_{DM}$--$\delta$ for $c^n/c^p$=-0.03 (the right--handed
  panel is a zoom--up of the left--handed one). The region bounded by
  the (red) dotted contour represents configurations where ${\cal
    D}<$1, while inside the (black) solid contour ${\cal
    D}_{Maxwellian}<$1. The horizontally (blue) hatched area is the
  same shown in Fig.\ref{fig:mchi_delta_na_f_i}(a), while the
  vertically (red) hatched region is the same shown in
  Fig.\ref{fig:mchi_delta_v_cut}. For the benchmark points $P_1$
  ((black) filled circle), $P_2$ ((green) square) and $P_3$ ((blue)
  star) the measurements and bounds for the functions $\tilde{\eta}$,
  $\tilde{\eta}_1$ used to calculate ${\cal D}$ and ${\cal
    D}_{Maxwellian}$ are explicitly plotted in
  Figs.\ref{fig:vmin_eta_benchmark_1},
  \ref{fig:vmin_eta_benchmark_2_3}(a) and
  \ref{fig:vmin_eta_benchmark_2_3}(b), respectively.}
\label{fig:mchi_delta_full_analysis}
\end{figure}

As already discussed in our Introduction and in Section
\ref{sec:fluorophobic}, the parameter $r=c^n/c^p$ must be chosen small
enough to evade the bounds from xenon detectors.  To illustrate this
point we provide in Fig.\ref{fig:r_d} a plot of ${\cal D}$ and ${\cal
  D}_{Maxwellian}$ as a function of the ratio $c^n/c^p$. In this
Figure, for the specific choice $m_{DM}$=11.4 GeV, $\delta$=23.7 keV
(chosen to be inside the horizontally (blue) hatched area of
Fig\ref{fig:mchi_delta_na_f_i}(a) and identified with the benchmark
$P_1$ in Fig. \ref{fig:mchi_delta_full_analysis}) the solid lines
represent ${\cal D}$ and ${\cal D}_{Maxwellian}$ where the
spin--dependent nuclear form factor for xenon is evaluated using Ref
\cite{haxton1,haxton2} (i.e. the same that we use also for all the
other nuclei) while (red) dots and (blue) dashes represent the same
quantities where we have used only for xenon two alternative
determinations of the spin--dependent form factor, Bonn-A and
Nijmegen, respectively, both taken from
Ref.\cite{spin_form_factors}. In each case the steeper curve is for
${\cal D}_{Maxwellian}$ while the shallower one represents ${\cal D}$,
and in order to have compatibility between DAMA and all other
constraints both compatibility factors must be below 1. The fact that
in Fig.\ref{fig:r_d} all curves change when only the form factor of
xenon is modified shows that, indeed, when the IDM parameters are
chosen as explained in Section \ref{sec:fluorophobic} the only
remaining bounds are from xenon targets (actually the compatibility
factor turns out to be driven by LUX which has a lower energy
threshold than XENON100). Moreover, the same Figure shows that in the
Maxwellian case the amount of the required cancellation between the
WIMP--proton and the WIMP--neutron amplitudes in xenon is
always larger than the hierarchy between the spin fractions carried by
protons and neutrons (i.e. ${\cal D}_{Maxwellian}>$1 for $c^n$=0 for
all three determination of the form factor). This means that for the
Maxwellian case the ratio $c^n/c^p$ must be tuned to a small but
non--vanishing value, which actually depends on which nuclear form
factor is adopted. In the following analysis we will use for all
nuclei the determination in \cite{haxton1,haxton2} and fix
$c^n/c^p$=-0.03 which suppresses the corresponding xenon
response\footnote{The corresponding values for the Bonn-A and Nijmegen
  form factors are $c^n/c^p\simeq$-0.08 and $c^n/c^p\simeq$-0.05,
  respectively, as can be seen in Fig.\ref{fig:r_d}.}.

The result of a systematic scanning of the two compatibility factors
${\cal D}$ and ${\cal D}_{Maxwellian}$ in the $m_{DM}$--$\delta$
parameter space is shown in Fig.\ref{fig:mchi_delta_full_analysis}. In
particular, the region enclosed in the (black) solid curve represents
the IDM parameter space where ${\cal D}_{Maxwellian}<1$, while the
parameter space inside the wider (red) dotted contour has ${\cal
  D}<1$. The regions which for kinematic reasons are not accessible to
fluorine and iodine targets (discussed in detail in Section
\ref{sec:fluorophobic}) are shown with (blue) horizontal hatches
(Maxwellian case) and with (red) vertical hatches
(halo--independent). They are the same as those shown in
Figs. \ref{fig:mchi_delta_na_f_i}(a) and
Fig. \ref{fig:mchi_delta_v_cut} and, as anticipated, cover a volume of
parameter space substantially smaller then the ones with ${\cal
  D}_{Maxwellian}<1$ and ${\cal D}<1$. This figure also also shows
that, as expected, in the halo--independent approach, corresponding to
the minimal set of assumptions for the WIMP velocity distribution, the
allowed parameter space is much wider compared to the Maxwellian case.

As far as the halo--independent analysis is concerned, by numerical
inspection we have determined that the lower part of the (red) dotted
boundary is due to the PICASSO constraint (driven by scatterings off
fluorine), while, when $m_{DM}\gsim$ 20 GeV, the upper part of the
boundary is due to COUPP (dominated by iodine). On the other hand,
when 13 GeV $\lsim m_{DM}\lsim$ 20 GeV the upper part of the dotted
boundary is determined by the shape test on the DAMA modulated
amplitudes, i.e. the condition $\Delta_{ST}>$1.96 becomes more
constraining than COUPP. Finally, in the upper part of the contour
with $m_{DM}\lsim$ 13 GeV the $v_{min}$ range explaining the DAMA
effect is driven beyond $v_{esc}^{lab}$, so $\Delta R_{2,4}<\Delta
R_{2,4,min}^{DAMA}$.  All these behaviours (with the exception of the
boundary driven by $\Delta_{ST}$) can be qualitatively understood in
terms of the kinematic boundaries discussed in
Fig.\ref{fig:mchi_delta_v_cut}.

\begin{figure}
\begin{center}
\includegraphics[width=0.49\columnwidth, bb=73 193 513 636]{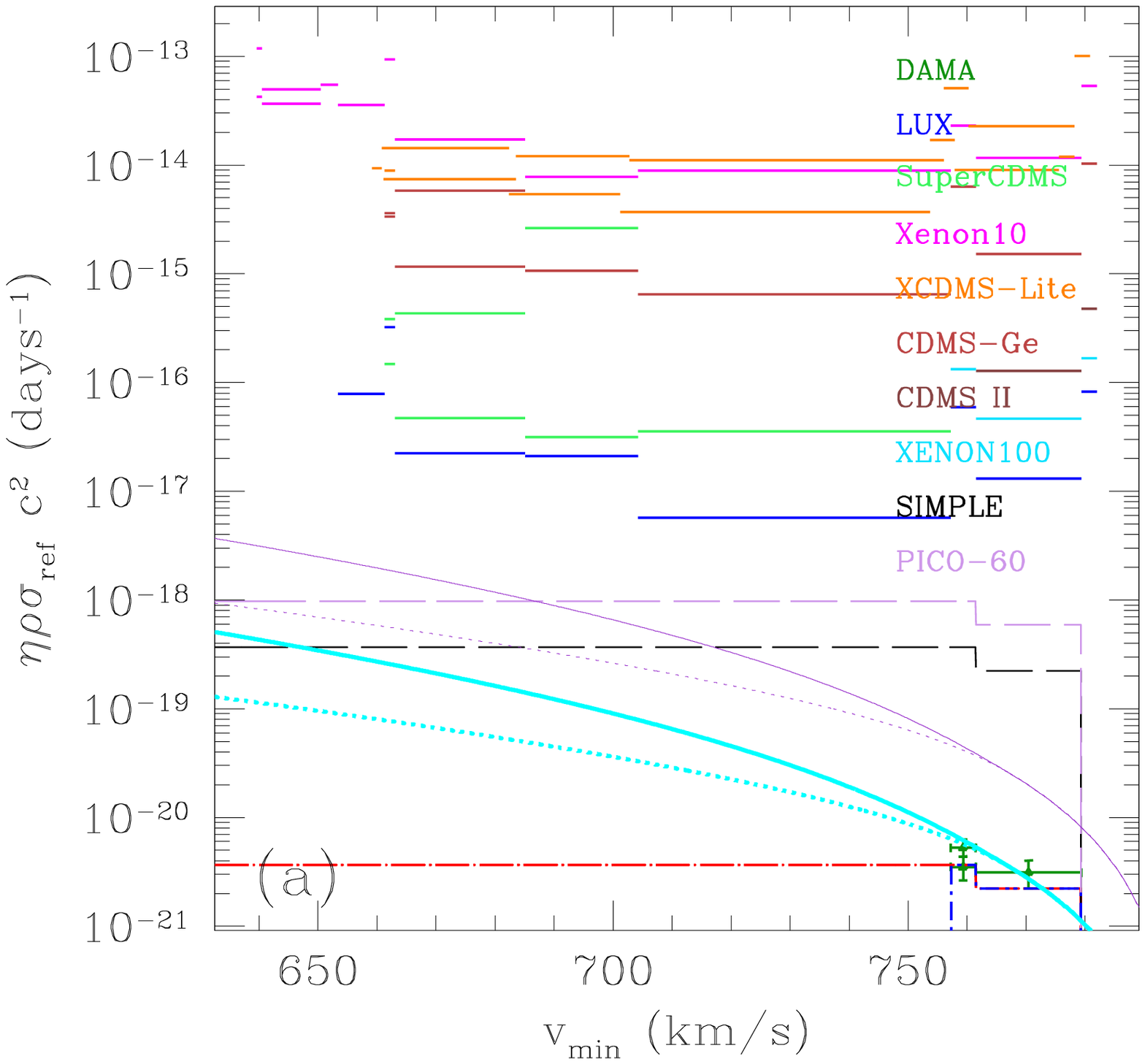}
\includegraphics[width=0.49\columnwidth, bb=73 193 513 636]{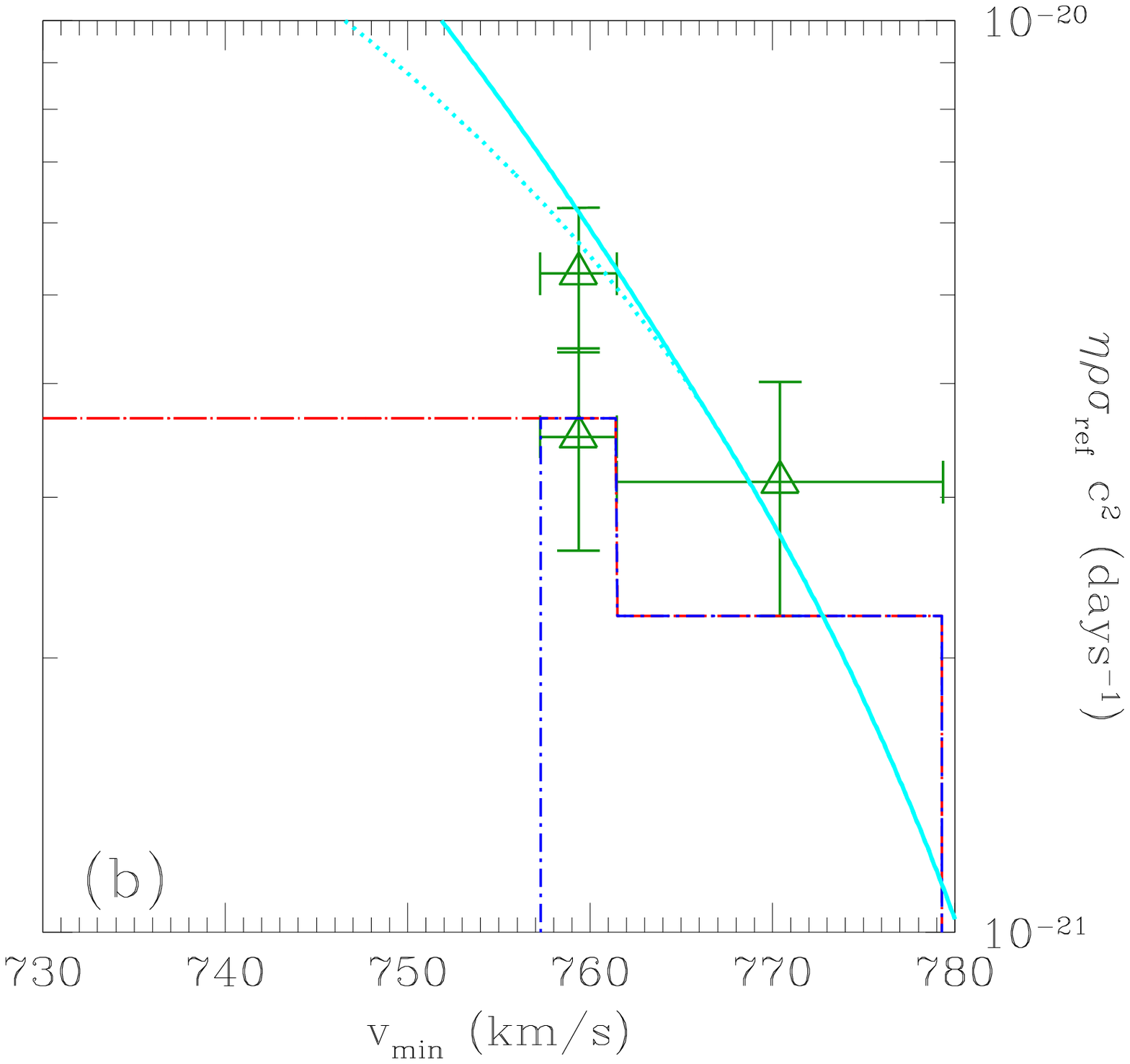}
\end{center}
\caption{{\bf (a)} Measurements and bounds for the functions
  $\tilde{\eta}$, $\tilde{\eta}_1$ defined in
  Eqs.(\protect\ref{eq:eta}) and (\protect\ref{eq:eta_modulation}) for
  $c^n/f^p$=-0.03 and $m_{DM}$=11.4 GeV, $\delta$=23.7 keV (the
  benchmark point $P_1$ indicated with a (black) filled circle in
  Fig.\ref{fig:mchi_delta_full_analysis}) used to calculate the
  compatibility factor ${\cal D}$
  (\ref{eq:compatibility_factor_eta_i}) and ${\cal D}_{Maxwellian}$
  (\ref{eq:compatibility_factor_maxwellian}). The (green) triangles
  represent the $\tilde{\eta}_1$ estimations from DAMA, while
  horizontal lines show upper bounds from other experiments, as
  indicated in the plot. The thick (cyan) dotted line represents the
  best--fit on DAMA points of the function $\tilde{\eta}_1$ in the
  case of a Maxwellian distribution with $v_{esc}$=550 km/sec and
  $v_{rms}$=270 km/sec, while the thick (cyan) solid line is the
  corresponding $\tilde{\eta}$. On the other hand the (purple) thin
  solid line represents the maximal $\tilde{\eta}$ for the Maxwellian
  case allowed by PICO-2L, while the (purple) thin dotted line is the
  corresponding $\tilde{\eta}_1$. The (blue) dots -- short dashes
  represent a conservative piecewise estimation of the function
  $\tilde{\eta}^{est}_1(v_{min})$ passing through DAMA points: (red)
  dots -- long dashes show the corresponding minimal piecewise
  estimation of the function $\tilde{\eta}^{est}(v_{min})$ (in
  compliance to the requirements of Eq.(\ref{eq:eta_conditions})) used
  to calculate the quantities $N_{k,i}^{expected}$ of
  Eq.(\ref{eq:compatibility_factor_generilized}) for droplet detectors
  and bubble chambers (see Section \ref{sec:compatibility}). The lower
  long--dashed (black) line shows the maximal
  $\tilde{\eta}^{est}(v_{min})$ allowed by scatterings off chlorine in
  SIMPLE, while the upper (purple) one is the same curve for
  scatterings off fluorine in PICO-60. {\bf (b)} A zoom-up of plot
  {\bf (a)} with the details of the piecewise functions
  $\tilde{\eta}^{est}_1(v_{min})$ and $\tilde{\eta}^{est}(v_{min})$.}
\label{fig:vmin_eta_benchmark_1}
\end{figure}

\begin{figure}
\begin{center}
\includegraphics[width=0.49\columnwidth, bb=73 193 513 636]{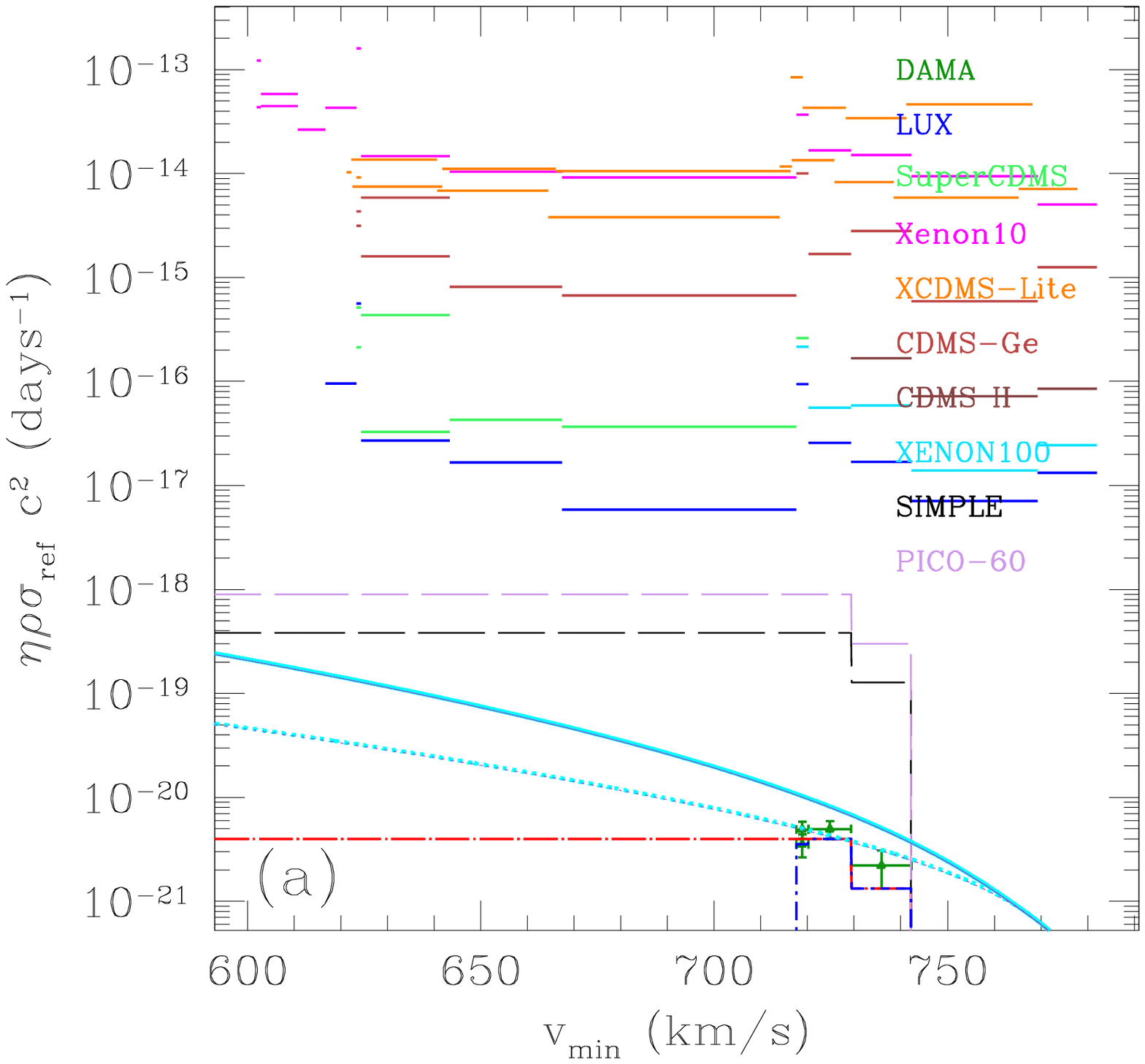}
\includegraphics[width=0.49\columnwidth, bb=73 193 513 636]{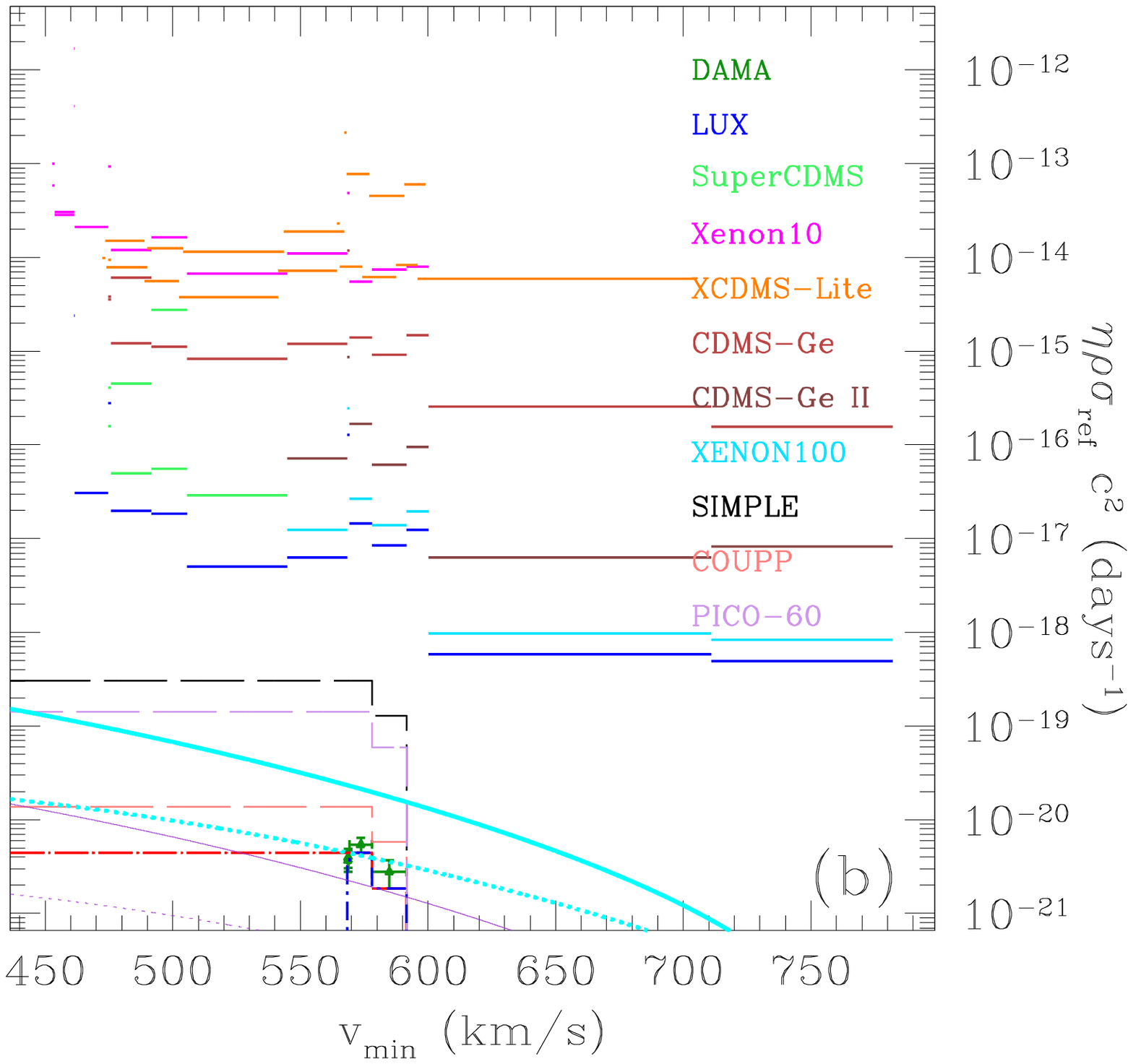}
\end{center}
\caption{{\bf (a)} The same as in Figure
  \ref{fig:vmin_eta_benchmark_1} with $m_{DM}$=12 GeV and $\delta$=22
  keV (the benchmark point shown as a (green) square in
  Fig.\ref{fig:mchi_delta_full_analysis}); {\bf (b)} The same as in
  Figure \ref{fig:vmin_eta_benchmark_1} with $m_{DM}$=17 GeV and
  $\delta$=17 keV (the benchmark point shown as a (blue) star in
  Fig.\ref{fig:mchi_delta_full_analysis}). }
\label{fig:vmin_eta_benchmark_2_3}
\end{figure}

In Fig.\ref{fig:mchi_delta_full_analysis} we have also selected three
benchmark points, indicated with a (black) filled circle
$P_1$=($m_{DM}$=11.4 GeV,$\delta$=23.7 keV), a filled (green) square
$P_2$=($m_{DM}$=12 GeV,$\delta$=22 keV) and a (blue) star
$P_3$=($m_{DM}$=17 GeV,$\delta$=17 keV), for which a detailed
discussion is provided in Figs.\ref{fig:vmin_eta_benchmark_1},
\ref{fig:vmin_eta_benchmark_2_3}(a) and
\ref{fig:vmin_eta_benchmark_2_3}(b), respectively.  The benchmark
$P_1$ lies inside the kinematic (blue) horizontally hatched area where
conditions
(\ref{eq:delta_bound_fluorine},\ref{eq:delta_bound_dama},\ref{eq:delta_bound_coupp})
apply. On the other hand, $P_2$ is outside such area but inside the
(red) vertically-hatched region where in
Eqs.(\ref{eq:delta_bound_fluorine},\ref{eq:delta_bound_dama},\ref{eq:delta_bound_coupp})
the conservative substitution $v_{esc}^{lab}\rightarrow
v_{cut}^{lab}$=$v^{lab}_{cut,DAMA}$ is made. Moreover, $P_2$ is at the
border of the ${\cal D}_{Maxwellian}<$1 region. Finally, $P_3$ is
representative of a situation where ${\cal D}_{Maxwellian}\gg$1 but
the more conservative halo--independent condition ${\cal D}<$1 is
satisfied.

In particular, in Figures \ref{fig:vmin_eta_benchmark_1} and
\ref{fig:vmin_eta_benchmark_2_3} the measurements and bounds of the
functions $\tilde{\eta}$, $\tilde{\eta}_1$ defined in
Eqs.(\protect\ref{eq:eta}) and (\protect\ref{eq:eta_modulation}) and
used to calculate the compatibility factors ${\cal D}$
(\ref{eq:compatibility_factor_eta_i}) and ${\cal D}_{Maxwellian}$
(\ref{eq:compatibility_factor_maxwellian}) for $P_1$, $P_2$ and $P_3$
are shown. In such figures the (green) triangles represent the
$\tilde{\eta}_1$ estimations from DAMA, were we used the modulation
amplitudes of Fig.6 of Ref.\cite{dama}, also reported in our
Fig. \ref{fig:e_ee_sm_dama} (the corresponding horizontal bars
represent the $v_{min}$ intervals mapped from the experimental ones on
$E^{\prime}$ while the vertical bars correspond to 1$\sigma$
fluctuations). On the other hand, in
Figs. \ref{fig:vmin_eta_benchmark_1}(a) and
\ref{fig:vmin_eta_benchmark_2_3} the other horizontal bars show upper
limits from calorimeters, ionizators and scintillators that directly
measure the nuclear recoil energy, and whose upper bounds can be
mapped into $v_{min}$ intervals (we include in our analysis
LUX\cite{lux}, XENON100\cite{xenon100}, XENON10\cite{xenon10},
CDMS-$Ge$\cite{cdms_ge}, CDMSlite \cite{cdms_lite},
SuperCDMS\cite{super_cdms} and CDMS II\cite{cdms_2015}). Moreover, in
the same figures we show with (blue) dots -- short dashes a piecewise
estimation of the minimal function $\tilde{\eta}^{est}_1(v_{min})$
passing through the DAMA points and with (red) dots -- long dashes the
ensuing minimal piecewise estimation of the function
$\tilde{\eta}^{est}(v_{min})$ obtained from
$\tilde{\eta}^{est}_1(v_{min})$ in compliance to the requirements of
Eq.(\ref{eq:eta_conditions}).  As explained in Section
\ref{sec:compatibility} we use $\tilde{\eta}^{est}(v_{min})$ to
calculate the expected count rates $N_{k,i}^{expected}$ used in
Eq.(\ref{eq:compatibility_factor_generilized}) to obtain the
compatibility factor ${\cal D}$ for experiments such as droplet
detectors and bubble chambers that only measure rates above a
threshold and that contain different target nuclei (implying that the
function $\tilde{\eta}(v_{min})$ cannot in general be directly
factorized and mapped into the same $v_{min}$ ranges of the DAMA
points): in Figs.  \ref{fig:vmin_eta_benchmark_1} and
\ref{fig:vmin_eta_benchmark_2_3} the long--dashed lines show the
maximal $\tilde{\eta}^{est}(v_{min})$ allowed by this class of
experiments when the corresponding constraints are applied (in our
analysis we include SIMPLE\cite{simple}, COUPP\cite{coupp},
PICASSO\cite{picasso}, PICO-2L\cite{pico2l} and PICO-60\cite{pico60}).

In Figure \ref{fig:vmin_eta_benchmark_1}(b) we provide a zoom--up of
Figure \ref{fig:vmin_eta_benchmark_1}(a) where the details of how the
piecewise functions $\tilde{\eta}^{est}_1(v_{min})$ and
$\tilde{\eta}^{est}(v_{min})$ are obtained. In this case the energy
$E_{ee}^{*Na}$=$q E_R^{*Na}\simeq$ 2.5 keVee falls in the range 2
keVee--4 keVee where the DAMA modulation amplitudes are detected and
we rebin the DAMA data so that $E_{ee}^{*Na}$ is one of the energy
boundaries and different energy bins map into same $v_{min}$
intervals. The result of this procedure is explicitly shown in terms
of the recoil energy in Fig.\ref{fig:e_ee_sm_dama}, which is
calculated for the same benchmark point $P_1$. In particular, in
Fig.\ref{fig:e_ee_sm_dama} the first and third energy bins map into
the first $v_{min}$ interval of Fig.\ref{fig:vmin_eta_benchmark_1}(b):
in this case we perform the shape test of Eq.(\ref{eq:shape_test}) on
the ensuing two determinations $\bar{\tilde{\eta}}_{1,1}$ and
$\bar{\tilde{\eta}}_{1,2}$ of the modulated halo function and if
$\Delta_{ST}<1.96$ the $\tilde{\eta}^{est}_1(v_{min})$ function is
taken as the lower range of the statistical combination of
$\bar{\tilde{\eta}}_{1,1}$ ad $\bar{\tilde{\eta}}_{1,2}$ assuming
Gaussian fluctuations.

Keeping in mind the discussion of Section \ref{sec:fluorophobic} it is
possible to understand why for the two benchmark points $P_1$ and
$P_2$ an interpretation of the DAMA modulation effect in terms of WIMP
inelastic scatterings is not constrained by any other experiment. In
the case of $P_1$, which lies in the (blue) horizontally hatched
contour of Fig.\ref{fig:mchi_delta_full_analysis}, WIMPs cannot
upscatter off fluorine because the required velocity is larger than
$v_{esc}^{lab}$, and also scatterings off iodine in COUPP are not
kinematically accessible for the same reason. On the other hand, in
the case of $P_2$, which lies inside the (red) vertically hatched
contour of Fig.\ref{fig:mchi_delta_full_analysis}, the required WIMP
incoming velocity for the same processes exceeds
$v_{cut,DAMA}^{lab}$. This implies that for both $P_1$ and $P_2$ the
only experiments sensitive to WIMPs besides DAMA and, to a much lesser
extent, xenon and germanium detectors, are SIMPLE (due to the presence
of chlorine) and PICO-2L (which contains iodine and has a threshold
lower than COUPP). The maximal $\tilde{\eta}^{est}(v_{min})$ functions
allowed by SIMPLE and PICO-2L, shown in
Figs.\ref{fig:vmin_eta_benchmark_1}(a) and
\ref{fig:vmin_eta_benchmark_2_3}(a) with the lower (black) and upper
(purple) long--dashed lines, respectively are well above the one which
explains the DAMA effect (shown with the (red) dots -- long
dashes)\footnote{The small response function of chlorine (for which we
  use the estimation of Appendix C in Ref. \cite{eft_spin}) and the
  very limited exposure collected by PICO--60\cite{pico60} below the
  threshold of COUPP imply that the sensitivities of these two
  experiments are never sufficient to directly probe the DAMA effect
  in our scenario. For this reason we have neglected them in the
  kinematic discussion of Section \ref{sec:fluorophobic}.}. On the
other hand, the benchmark point $P_3$ is kinematically accessible to
both fluorine and iodine in COUPP: the corresponding maximal
$\tilde{\eta}^{est}(v_{min})$ function allowed by COUPP exceeds the
one required to explain DAMA and is shown in Figure
\ref{fig:vmin_eta_benchmark_2_3}(b) with the lowest (crimson)
long--dashed curve. In this case, as in all the outer region bounded
by the (red) dotted contour in Fig.\ref{fig:mchi_delta_full_analysis},
simple kinematic arguments cannot guarantee that a configuration is
allowed: also dynamical aspects (such as the cross--section scaling
laws among sodium, fluorine and iodine) and experimental issues (such
as the collected data exposures and cut acceptances) are needed to
ascertain the compatibility of DAMA with other experimental limits,
and a full calculation of the compatibility factor ${\cal D}$
(\ref{eq:compatibility_factor_generilized}) is required.

As far as the Maxwellian analysis is concerned, the lower part of the
solid (black) boundary in Fig. \ref{fig:mchi_delta_full_analysis} is
determined by fluorine targets, namely PICO-2L for $_{DM}\gsim$11 GeV
and PICO-2L+PICASSO for $_{DM}\lsim$11 GeV. On the other hand, the
right-hand side of the contour is driven by WIMP--iodine scatterings
in COUPP. Finally, in the remaining left and upper parts of the
contour our scenario cannot give a satisfactory interpretation of the
DAMA effect because the $p$--value of the minimal $\chi$--square of
the DAMA modulation amplitudes corresponding to the best estimation
$(\rho_{DM}\sigma_{ref})_{best}$ of $\rho_{DM}\sigma_{ref}$ is below
0.05.

In Figures \ref{fig:vmin_eta_benchmark_1} and
\ref{fig:vmin_eta_benchmark_2_3} $(\rho_{DM}\sigma_{ref})_{best}$ is
used to calculate the functions $\tilde{\eta}^{est}(v_{min})$ and
$\tilde{\eta}^{est}_1(v_{min})$ in the Maxwellian case: the
corresponding curves are shown with the thick (cyan) solid and dotted
curves, respectively. On the other hand, in the same Figures the thin
(purple) solid and dotted curves represent the maximal
$\tilde{\eta}^{est}(v_{min})$ and $\tilde{\eta}^{est}_1(v_{min})$
allowed by the most constraining among the other experiments, i.e. the
two functions are calculated using the smallest among the cross
section upper limits $\sigma_{ki}^{bound}$ used in the compatibility
factor ${\cal D}_{Maxwellian}$ of
Eq.(\ref{eq:compatibility_factor_maxwellian}) for each experiment $k$
and each energy bin $i$.

Also for the Maxwellian case the considerations of Section
\ref{sec:fluorophobic} can be helpful in interpreting the numerical
results of Figs \ref{fig:vmin_eta_benchmark_1} and
\ref{fig:vmin_eta_benchmark_2_3}.  In particular, it is clear from our
previous considerations that the benchmark point $P_1$ is not
accessible to fluorine detectors and to iodine in COUPP: indeed, in
Figure \ref{fig:vmin_eta_benchmark_1} the $\sigma_{ki}^{bound}$ used
to calculate the maximal $\tilde{\eta}^{est}(v_{min})$ and
$\tilde{\eta}^{est}_1(v_{min})$ functions is from chlorine in SIMPLE
and is well above the corresponding function required to explain the
DAMA effect. On the other hand, the benchmark point $P_2$ was chosen
close to the (black) solid contour of
Fig. \ref{fig:mchi_delta_full_analysis} where ${\cal
  D}_{Maxwellian}\simeq$1: for this reason in
Fig.\ref{fig:vmin_eta_benchmark_2_3}(a) the
$\tilde{\eta}^{est}(v_{min})$ function calculated using
$(\rho_{DM}\sigma_{ref})_{best}$ is very close to the maximal Maxwellian
halo function allowed by other experiments. Finally, the benchmark
point $P_3$ is well outside the ${\cal D}_{Maxwellian}<$1 parameter
space, so in Figure \ref{fig:vmin_eta_benchmark_2_3}(b) the best--fit
estimation of the Maxwellian halo function is much above the maximal
$\tilde{\eta}^{est}(v_{min})$ allowed by other constraints. For both
$P_2$ and $P_3$ the most constraining $\sigma_{ki}^{bound}$ is from
PICO-2L.

\begin{figure}
\begin{center}
\includegraphics[width=0.49\columnwidth, bb=73 193 513 636]{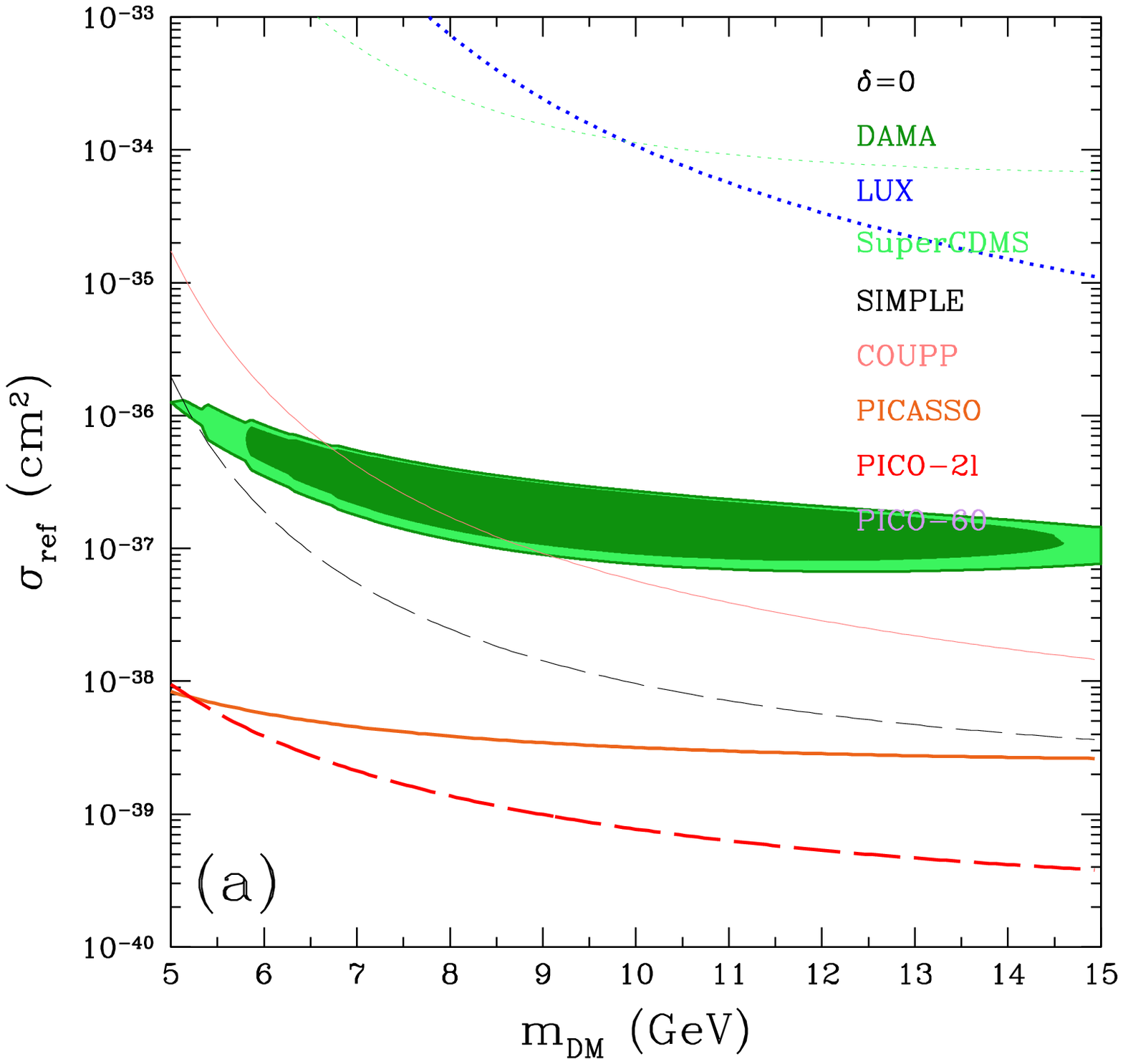}
\includegraphics[width=0.49\columnwidth, bb=73 193 513 636]{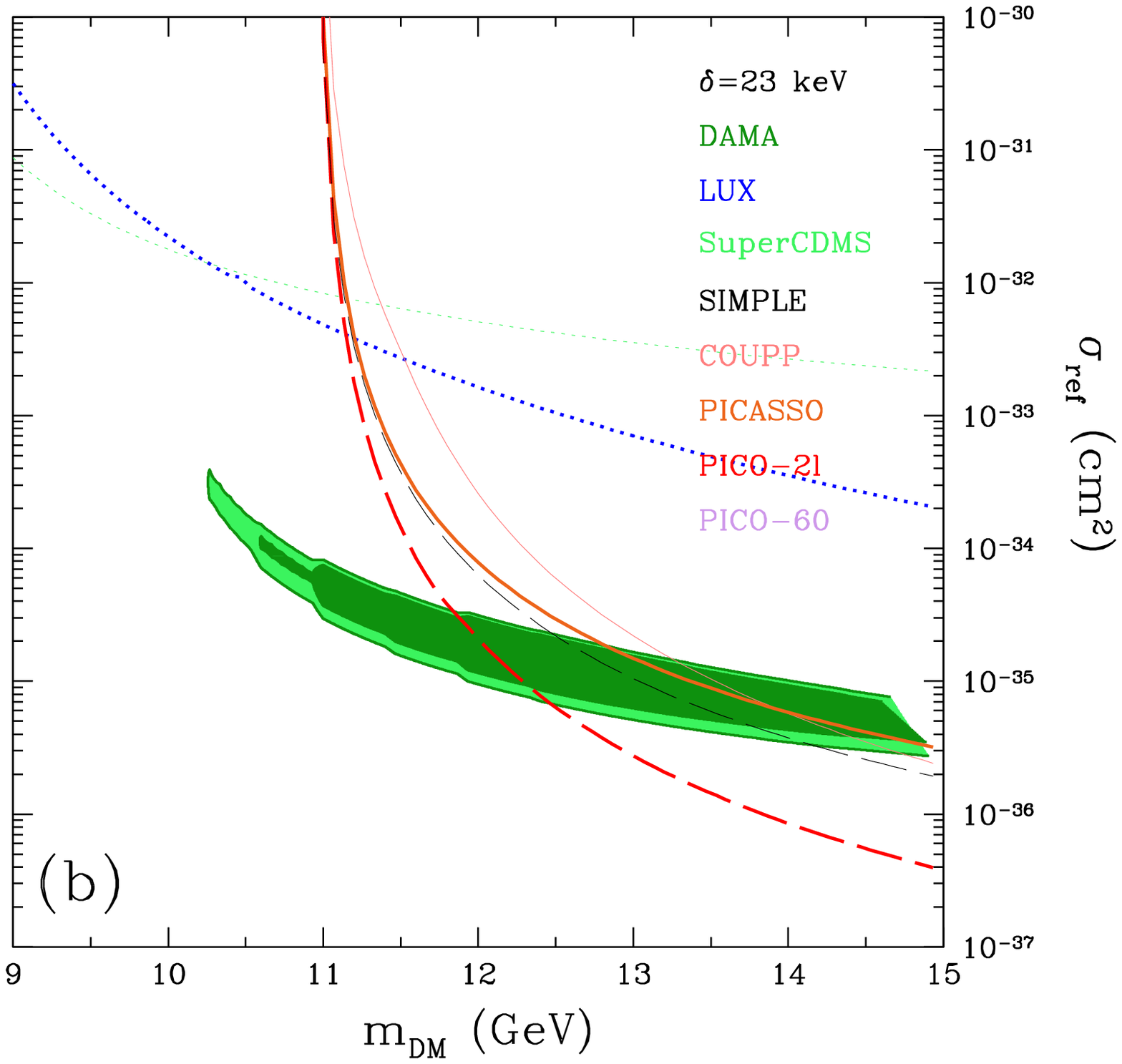}
\end{center}
\caption{{\bf (a)} Light and dark (green) shaded areas represent the
  99 \% C.L. and 95 \% C.L. parameter space ranges compatible to the
  DAMA modulation effect in the $m_{DM}$--$\sigma_{ref}$ plane for
  $\delta$=0. Open curves show the corresponding 90\% C.L. upper
  bounds from other experiments: LUX\cite{lux} (thick (blue) dots),
  SuperCDMS\cite{super_cdms} (thin (green) dots), SIMPLE\cite{simple}
  (thin (black) dashes), COUPP\cite{coupp} (thin (red) solid line),
  PICASSO\cite{picasso} (thick (orange) solid line) and
  PICO--2L\cite{pico2l} (thick (red) dashes).  {\bf (b)} The same for
  $\delta$=23 keV. In both plots $c^n$/$c^p$=-0.03 and $\rho_{DM}$=0.3
  GeV/cm$^3$.}
\label{fig:mchi_sigma_eft_inelastic}
\end{figure}

The identification of $f(\vec{v})$ with a Maxwellian distribution
allows the factorization of the reference cross section $\sigma_{ref}$
as a function of the WIMP mass $m_{DM}$ (at fixed $\delta$ for
inelastic scattering). Indeed, this is the standard procedure adopted
by experimental collaborations to present their data. For
$\rho_{DM}$=0.3 GeV/cm$^3$ we show the result of such analysis when
$\delta$=0 (elastic scattering case) in
Fig.\ref{fig:mchi_sigma_eft_inelastic}(a) while in
Fig.\ref{fig:mchi_sigma_eft_inelastic}(b) we adopt $\delta$=23 keV. In
both figures the light and dark (green) shaded areas represent the 99
\% C.L. and the 95 \% C.L. $m_{DM}$--$\sigma_{ref}$ parameter space
compatible to the DAMA modulation effect while open curves show the
corresponding 90\% C.L. upper bounds on $\sigma_{ref}$ from other
experiments. As in all the other plots here $c^n/c^p$=-0.03. Indeed,
in the elastic case ($\delta$=0) of
Fig.\ref{fig:mchi_sigma_eft_inelastic}(a) the constraints from bubble
chambers and droplet detectors, all containing fluorine, are in strong
tension to an interpretation of the DAMA modulation effect in terms of
a WIMP signal. However, as shown in
Fig.\ref{fig:mchi_sigma_eft_inelastic}(b) for $\delta$=23 keV, when
$m_{DM}$ and $\delta$ fall inside the solid (black) contour of
Fig.\ref{fig:mchi_delta_full_analysis} all fluorine upper bounds are
kinematically relaxed.

We conclude this Section by noting that, when $f(\vec{v})$ is
identified with a Maxwellian distribution, in Figures
\ref{fig:vmin_eta_benchmark_1} and \ref{fig:vmin_eta_benchmark_2_3}(a)
the $\tilde{\eta}^{est}_1(v_{min})$ modulated halo function traces
very closely $\tilde{\eta}^{est}(v_{min})$ if $v_{min}$ is in the
range of the DAMA effect, i.e. the modulation amplitude fraction
predicted in DAMA is very close to unity. This is at variance with the
qualitative expectation that the modulated fraction is of order
$\Delta v_{Earth}$/$v_{Earth}\simeq$0.07. For $P_1$ the effect is
quite dramatic although, as discussed in detail in Section
\ref{sec:fluorophobic}, the two parameters $m_{DM}$ and $\delta$ must
be considerably tuned to verify the condition of
Eq.(\ref{eq:hierarchy}) and are sensitive to the choice of
$v_{esc}$. However, a large modulation fraction is also present inside
the wider ${\cal D}_{Maxwellian}<$1 contour: indeed, as shown in
Fig. \ref{fig:vmin_eta_benchmark_2_3}(a), the benchmark point $P_2$,
which is on the boundary of such region of the parameter space, has
modulation fractions of order 50\% in the DAMA $v_{min}$ range. For
this effect to happen it is clearly necessary that a combination of
small $m_{DM}$ and large $\delta$ drive $v_{min}$ in DAMA to the
large--velocity tail of $f(\vec{v})$. On the other hand, when the IDM
parameter space is enlarged to the wider (red) dotted contour of
Fig. \ref{fig:mchi_delta_full_analysis} $v_{min}$ moves to lower
values where such large modulation fractions may not be necessarily
expected. Indeed in Fig. \ref{fig:vmin_eta_benchmark_2_3}(b) for the
Maxwellian case (which is anyway excluded by the combined constraints
from PICASSO, PICO-60 and PICO-2L)
$\tilde{\eta}^{est}_1(v_{min})$$\ll$$\tilde{\eta}^{est}(v_{min})$. On
the other hand when the the halo--independent analysis is applied
$\tilde{\eta}^{est}(v_{min})\simeq\tilde{\eta}^{est}_1(v_{min})$ is
going to be eventually required at the boundary of the (red) dotted
contour of Fig. \ref{fig:mchi_delta_full_analysis} to evade the
constraints on DAMA from other experiments, although deep in the
allowed region smaller modulation amplitudes are allowed: for example,
in Fig. \ref{fig:vmin_eta_benchmark_2_3}(b) the halo--independent
$\tilde{\eta}^{est}_1(v_{min})$ from DAMA is about 33\% of the
corresponding maximal $\tilde{\eta}^{est}(v_{min})$ allowed by the
most stringent constraint (COUPP) represented by the lowest (crimson)
long--dashed line.

When present, the details of the large--modulation effect in our
scenario are expected to depend on the specific choice of the velocity
distribution. In particular, the large--speed regime of $f(\vec{v})$
may correspond to sparsely populated regions in the WIMP phase space
less likely to be well predicted by the gross approximation of a
Maxwellian with a velocity cut--off: indeed, they may contain
non--thermalized components difficult to predict in numerical
simulations and dependent on the merger history of our Galaxy
\cite{non_thermal}. Nevertheless a large-velocity cut--off in
$f(\vec{v})$ and a steep dependence of the halo function in the large
$v_{min}$ regime are features expected on quite general grounds, so
that large modulated fractions should be considered a natural
prediction of our scenario. In the following Section we will use a
simple Maxwellian distribution cut--off at $v_{esc}$ to capture at
least qualitatively such features in order to address the issue of
whether, if possible, such modulated fractions may be constrained or
even already excluded in the DAMA data.

We close this Section observing that modulation fractions much larger
than usually expected might be an attractive possibility when
explaining DAMA. In fact a theoretical prediction of the modulation
fraction $\Delta R_{[\Ed_1, \Ed_2]}$/$R_{[\Ed_1, \Ed_2]}$ allows to
use the DAMA measurement of $\Delta R_{[\Ed_1, \Ed_2]}$ to get an
estimation of the $R_{[\Ed_1, \Ed_2]}$ contribution of the signal to
the time--averaged event spectrum measured by DAMA, and so, by
difference, of the background contribution from radioactive
contamination, neutrons and cosmic rays. The fact that the low--energy
time--averaged count rate measured by DAMA is approximately flat,
while instead the predicted $R_{[\Ed_1, \Ed_2]}$ usually depends
exponentially on the recoil energy, has been used by some authors to
claim that in some cases the resulting estimation of the background
shows an energy dependence difficult to reconcile to what is expected
by simulations \cite{dama_bck}. Probably the many uncertainties of the
latter do not allow to draw robust constraints anyway: however, large
modulation fractions imply much smaller predictions for $R_{[\Ed_1,
  \Ed_2]}$ easing this potential tension.

\section{Large modulation fractions: the Maxwellian case}
\label{sec:modulation}

In the previous Sections we pointed out that a natural prediction of
our scenario is that the annual time-variation of the rate of WIMP--Na
scatterings in DAMA can be much larger than usually expected and even
approaching 100\% of the time-averaged rate.  In particular, this is
expected to happen on quite general grounds if the kinematic condition
(\ref{eq:hierarchy}) is verified, and can be verified quantitatively
in the case of a Maxwellian WIMP velocity distribution.

It has been claimed in the literature that in such case a significant
distortion of the time dependence of the expected rate from a cosine
should be present\cite{freese}. In this Section we wish to elaborate
more on this aspect. In order to be quantitative, we will explicitly
adopt for $f(\vec{v})$ a Maxwellian distribution defined in the
reference frame of the Galaxy with a cut for $|\vec{v}|<v_{esc}$, with
the reference values $v_{rms}\simeq$ 270 km/sec, $v_{esc}$=550 km/sec.
Moreover, for definiteness in this Section we will adopt the set of
IDM parameters $m_{DM}$=11.4 GeV and $\delta$=23.7 keV, which
corresponds to the benchmark point $P_1$ in
Fig.\ref{fig:mchi_delta_full_analysis} and maximizes the effect.

\begin{figure}
\begin{center}
\includegraphics[width=0.49\columnwidth, bb=73 193 513 636]{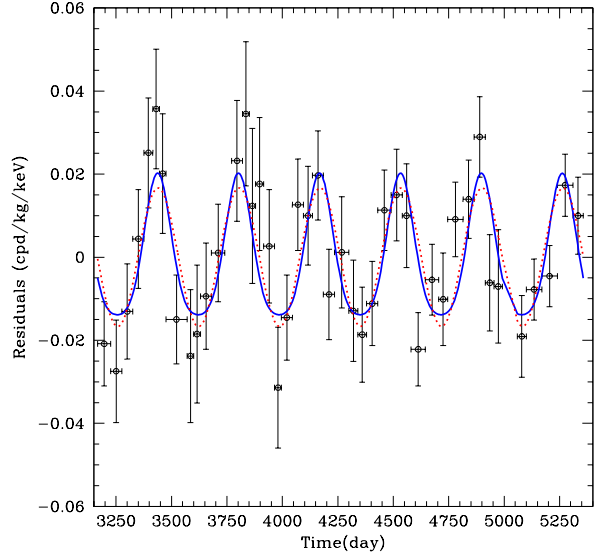}
\end{center}
\caption{The data points represent the count--rate residuals measured
  by DAMA in the energy interval 2--4 keVee using an exposure of
  0.87 ton$\times$ year (from Fig. 1 of \cite{dama}). Solid (blue)
  line: theoretical prediction of the count--rate residuals for the
  benchmark point $P_1$=($m_{DM}$=11.4 GeV,$\delta$=23.7 keV) assuming
  a Maxwellian velocity distribution when the product $\rho_{DM}
  \sigma_{ref}$ is set to the value which minimizes a $\chi$--square
  with the 43 experimental data points; dotted (red) line: the same
  quantity for a cosine time dependence. For both theoretical
  predictions we have assumed a phase of the modulation corresponding
  to $t_0$=152.5 days (June 2nd).}
\label{fig:dama_residuals_2_4_kev}
\end{figure}

\begin{figure}
\begin{center}
\includegraphics[width=0.49\columnwidth, bb=73 193 513 636]{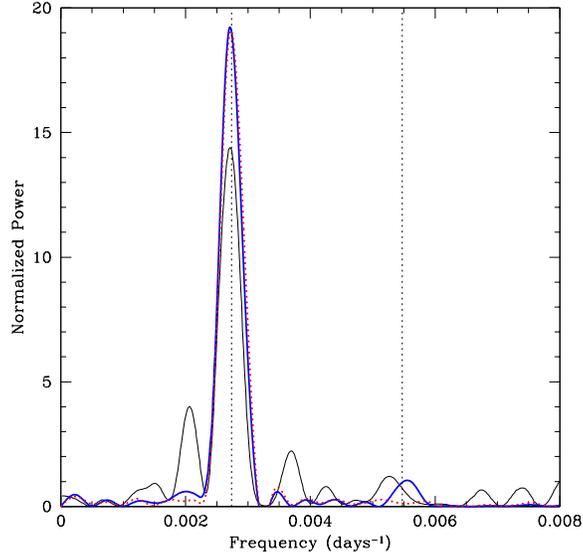}
\end{center}
\caption{Thin (black) solid line: experimental DAMA determination of
  the normalized power spectrum in the energy interval 2 keVee$ \le
  E_{ee}\le$ 6 keVee and using an exposure of 0.87 ton$\times$ year
  (from Fig.2 of \cite{dama}); thick solid (blue) line: prediction of
  the same quantity for the $P_1$ benchmark; thick (red) dotted line:
  the same for a cosine time dependence.}
\label{fig:power_spectrum}
\end{figure}

In Figs.\ref{fig:dama_residuals_2_4_kev} and \ref{fig:power_spectrum}
we show the annual modulation effect in time and frequency space
comparing the predictions for benchmark $P_1$ to the DAMA data
\cite{dama}. In particular, in Fig. \ref{fig:dama_residuals_2_4_kev}
the data points are taken from Fig. 1 of \cite{dama} and represent the
count--rate residuals measured by DAMA in the energy interval 2--4
keVee using an exposure of 0.87 ton$\times$ year (corresponding to a
total period of 2191 days and encompassing 6 complete annual
modulation cycles). Since the full DAMA data are not available to
analyze, we will use these data points for a comparison with our
large--modulation scenario. In each data point the horizontal error
bar represents the time intervals over which the detected count rate
has been averaged (the data are averaged over 43 time intervals whose
amplitude ranges from $\simeq$30 to $\simeq$70 days and which are
shorter close to maxima and minima). The residuals are obtained by
averaging the measured count rate in each time interval (after cuts to
reduce the background) and subtracting the total average, which is
obtained by dividing the total event sum over the complete period of
data taking by the number of live days. In the same Figure the solid
(blue) line represents the $P_1$ theoretical prediction calculated in
the same way. This line has been calculated by fixing the product
$\rho_{DM} \sigma_{ref}$ to the value which minimizes a $\chi$--square
with the 43 experimental data points. In the same figure the dotted
(red) line represents the same quantity obtained by assuming for the
residual a cosine time dependence. In both cases we have assumed a
phase of the modulation corresponding to $t_0$=152.5 days (June
2nd). Indeed, in the $P_1$ prediction a distortion is clearly
visible. However, the minimal $\chi$--square in the two cases is 40.8
and 42.3 respectively, with 43-1 degrees of freedom: within the
experimental errors the two time dependencies are undistinguishable.

The distortion visible in the time dependence of the model $P_1$
expected rate can be traced back to the fact that the latter can be
written as a power expansion in terms of the small parameter $\Delta
v_{Earth}/v_{Sun} \cos[2\pi t/T-t_0]\simeq 0.07 \cos[2\pi t/T-t_0]$:

\begin{eqnarray}
  S[t]&=&S_0+S_1 \frac{\Delta v_{Earth}}{v_{Sun}}\cos \left [\frac{2\pi}{T} t-t_0 \right ]+S_2 \left ( \frac{\Delta v_{Earth}}{v_{Sun}}\cos \left [\frac{2\pi}{T} t-t_0 \right ] \right )^2+... \nonumber \\
  &=&S_0+\tilde{S}_1 \cos \left [\frac{2\pi}{T} t-t_0 \right ]+\tilde{S}_2 \left (\cos \left [\frac{2\pi}{T} t-t_0 \right ] \right )^2 +...
\end{eqnarray}

\noindent In the case of large modulation fractions one has $S_0\simeq
\tilde{S}_1$ and when $\cos[2\pi/T t-t_0]\simeq$-1, the cancellation
between the first two terms in the power expansion implies dominance
of $\tilde{S}_2$, explaining the distortion in
Fig.\ref{fig:dama_residuals_2_4_kev}. Nevertheless, due to the small
ratio $\Delta v_{Earth}/v_{Sun}$ the term $\tilde{S}_2$ is always much
smaller than $S_0$ and $\tilde{S}_1$. In particular this implies that
also in frequency space the contribution of the second harmonics
remains necessarily small. This can be seen in
Fig. \ref{fig:power_spectrum}, where the thin (black) solid line
represents the experimental DAMA determination of the normalized power
spectrum in the energy interval 2 keVee$ \le E^{\prime}\le$ 6 keVee
(taken from Fig.2 of \cite{dama}), while the thick solid (blue) line
and the thick (red) dotted line show the predictions for the same
quantity using the expected rate for $P_1$ (calculated with the same
procedure of Fig.\ref{fig:dama_residuals_2_4_kev}) and a cosine time
dependence, respectively: inspection of this figure confirms that the
DAMA data are in substantial agreement with both the cosine dependence
and the distorted one. So we conclude that, since the distortions in
the time dependence of the annual modulation that arise in a natural
way in our scenario are compatible to the DAMA data even for the
extreme $\simeq$100\% case of benchmark $P_1$, this is also true in
the remaining part of the parameter space, where this effect is
expected to be milder.

\begin{figure}
\begin{center}
\includegraphics[width=0.49\columnwidth, bb=73 193 513 636]{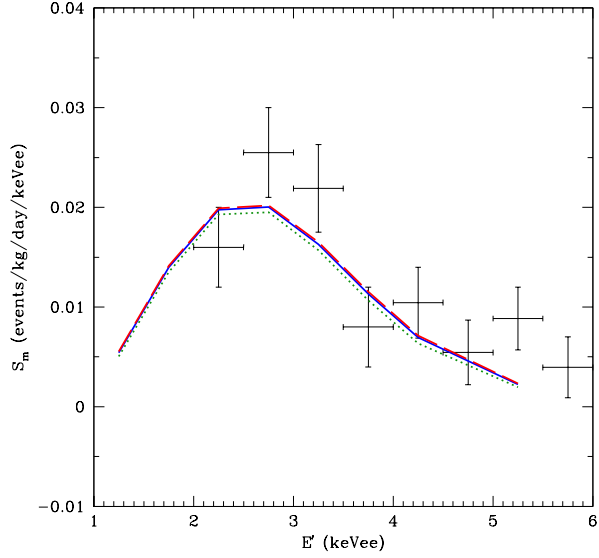}
\end{center}
\caption{The data points show the DAMA modulation amplitudes as a
  function of the detected recoil energy E$^{\prime}$ (from Fig. 6 of
  Ref.\cite{dama}). The other curves show theoretical predictions of
  the modulation amplitudes $S_m^{th}$ for the benchmark point $P_1$
  using different definitions (see text): (red) long dashes:
  $S_m^{th}$=$S_m^{diff}(E^{\prime})\equiv[S(E^{\prime},t_{max})-S(E^{\prime},t_{min})]/2$;
  (blue) solid line: $S_m^{th}$=$S_m^{\chi}(E^{\prime})$ minimizing
  the function $\chi^2=\sum_t \left
    [S(E^{\prime},t)-S_0^{\chi}(E^{\prime})-S_m^{\chi}(E^{\prime})
    \cos (2\pi t/T-t_0)\right ]^2$ with $S(E^{\prime},t)$ the full
  calculation of the expected rate; (green) dotted line:
  $S_m^{th}$=$S_m^{\delta}(E^{\prime})\equiv\partial
  S(E^{\prime},t)/\partial \eta\times \Delta \eta$ with
  $\eta=v_{Earth}/v_{Sun}$ and $\Delta \eta=\Delta
  v_{Earth}/v_{Sun}$.}
\label{fig:e_ee_sm_dama}
\end{figure}

We close this Section with a discussion of the energy dependence of
the modulation amplitudes in our scenario.  In
Fig.\ref{fig:e_ee_sm_dama} the data points represent the determination
of such quantities by DAMA.  The procedure adopted to extract them
from the experimental count rates is described in Ref.\cite{dama}: it
consists in parameterizing the expected rate in terms of the sum of a
constant part and a cosine--modulated one, i.e.
$S(E^{\prime},t)=B(E^{\prime})+S_0(E^{\prime})+S_m(E^{\prime}) \cos
(2\pi t/T-t_0)$ (where $B(E^{\prime})$ represents a time--independent
unknown background) and minimizing a likelyhood function with the data
in terms of $B(E^{\prime})$, $S_0(E^{\prime})$ and $S_m(E^{\prime})$,
binning the data in $\Delta E^{\prime}$=$q \Delta E_R=0.5$ keVee
intervals starting from 2 keVee (as can be seen from
Fig.\ref{fig:e_ee_sm_dama} the bulk of the modulation excess is
concentrated for $E^{\prime}<$4 keVee).  The experimental
determinations $S^{exp}_m$ obtained in this way for the modulation
amplitudes as a function of the recoil energy are then compared to the
theoretical predictions $S_m^{th}$ for a specific DM model.

However, when the time dependence of the count rate of a given DM
candidate departs from a cosine the $S_m^{th}$ definition deserves
some care.  In Fig. \ref{fig:e_ee_sm_dama} we show different
definitions of $S_m^{th}$ for benchmark $P_1$: the (red) long dashes
represent the theoretical prediction of the difference
$S_m^{diff}(E^{\prime})=[S(E^{\prime},t_{max})-S(E^{\prime},t_{min})]/2$,
while the (blue) solid line represents the quantity
$S_m^{\chi}(E^{\prime})$ that, together with $S_0^{\chi}(E^{\prime})$,
minimizes the function
$\chi^2[S_0^{\chi}(E^{\prime}),S_m^{\chi}(E^{\prime})]=\sum_t \left
  [S(E^{\prime},t)-S_0^{\chi}(E^{\prime})-S_m^{\chi}(E^{\prime}) \cos
  (2\pi t/T-t_0)\right ]^2$ where $S(E^{\prime},t)$ is the full
calculation of the expected rate for $P_1$ as a function of time
(notice that in this way $S_m^{\chi}/S_0^{\chi}>$1 is possible,
because the time--dependence of the rate is parameterized in terms of
a "wrong" functional form: indeed, in the particular example of Figure
\ref{fig:e_ee_sm_dama} this ratio is slightly higher than unity in all
the energy range\footnote{If the function $\chi^2$ is minimized
  imposing the condition $S_m^{\chi}/S_0^{\chi}\le$1 the modulated
  fraction of the signal turns out to be exactly 100\%.}).  Moreover,
in the same Figure the (green) dotted line represents the quantity
$S_m^{\delta}(E^{\prime})\equiv\partial S(E^{\prime},t)/\partial
\eta\times \Delta \eta$ with $\eta=v_{Earth}/v_{Sun}$ and $\Delta
\eta=\Delta v_{Earth}/v_{Sun}$. In case of a cosine time dependence of
$S(E^{\prime},t)$ the three definitions for the modulation amplitudes
$S_m^{diff}$, $S_m^{\chi}$ and $S_m^{\delta}$ coincide, but when the
time dependence departs from a simple cosine the $S_m^{th}$'s whose
definition is closer to the experimental $S_m^{exp}$ amplitudes
published by DAMA should be identified with the $S_m^{\chi}$'s
(lacking a full minimization of the likelihood function of the
data). Nevertheless, as can be seen in Fig. \ref{fig:e_ee_sm_dama},
the three determinations are very close to each other and their
difference much smaller than the experimental errors.  This justifies
the use of the definition $S_m^{th}=S_m^{diff}$ for the theoretical
predictions $S_m^{th}$ of the modulated amplitudes that was implied by
Eq.(\ref{eq:eta_modulation}) and used in Section \ref{sec:analysis}.

\begin{figure}
\begin{center}
\includegraphics[width=0.49\columnwidth, bb=73 193 513 636]{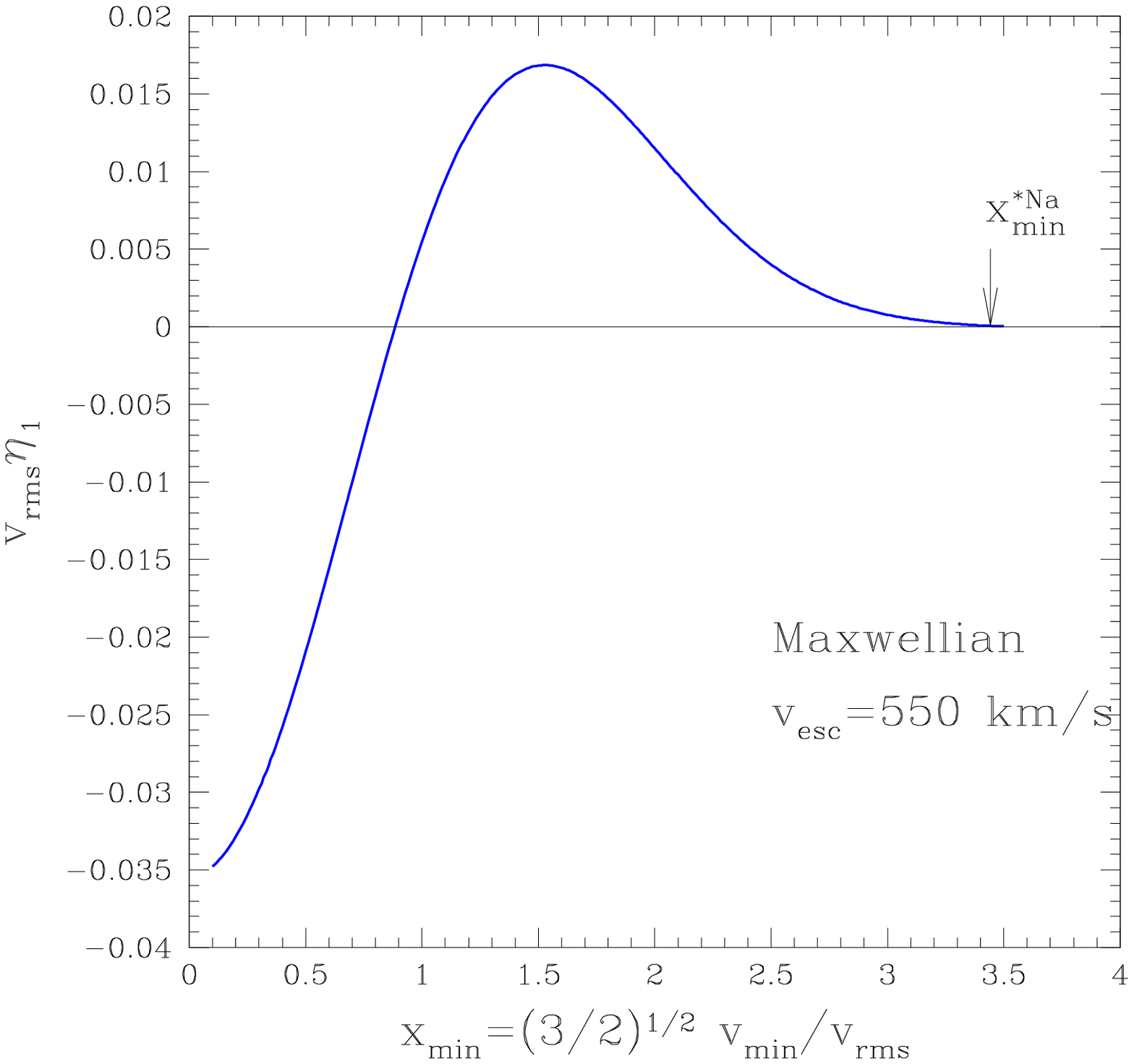}
\end{center}
\caption{Modulated halo function $\eta_1$ (corresponding to the
  $\tilde{\eta}_1$ of Eq.(\ref{eq:eta_modulation}) with
  $\rho_{DM}\sigma_{ref}$/$m_{DM}$$\rightarrow$1) as a function of the
  non-dimensional ratio $x_{min}\equiv \sqrt{3/2}v_{min}/v_{rms}$ for
  a Maxwellian distribution with $v_{rms}$=270 km/sec and $v_{esc}$=550
  km/sec. The arrow shows the position of
  $x_{min}^{*Na}$=$\sqrt{3/2}v_{min}^{*Na}/v_{rms}$ for the benchmark
  point $P_1$, i.e. for $m_{DM}$=11.4 GeV and $\delta$=23.7 keV.}
\label{fig:xmin_eta1}
\end{figure}

We conclude this Section by noting that, as anticipated, in
Fig.\ref{fig:e_ee_sm_dama} the function $S_m^{th}[E^{\prime}]$ shows a
maximum when 2 keVee$\lsim E^{\prime}\lsim$ 3 keVee, which is in rough
agreement to the experimental determinations $S_m^{exp}[E^{\prime}]$.
As shown in Fig.\ref{fig:xmin_eta1}, where the Maxwellian $\eta_1$
(corresponding to the modulated halo function $\tilde{\eta}_1$
(\ref{eq:eta_modulation}) with
$\rho_{DM}\sigma_{ref}$/$m_{DM}$$\rightarrow$1) is plotted as a
function of the non--dimensional combination $x_{min}\equiv
\sqrt{3/2}v_{min}/v_{rms}$, for elastic scattering this may be
interpreted as a possible indication that the WIMP mass is
particularly light ($m_{DM}\lsim$ 10 GeV), since this corresponds to
the low $x_{min}$ regime where the modulation amplitude shows a phase
inversion just below the DAMA energy
threshold\cite{inverse_modulation}\footnote{The presence of a peak in
  the measured modulation amplitudes is the reason why the DAMA data,
  while showing for elastic scattering two local mass minima of the
  Likelyhood, have a preference for the low--mass
  solution.}. Interestingly, in the case of our inelastic scattering
scenario the expected modulation amplitude $S_m^{th}[E^{\prime}]$
shows a maximum within the DAMA energy range for a completely
different reason. In fact in this case large $x_{min}$ values are
involved, in particular beyond the maximum of $\eta_1$, in a regime
where the latter is decreasing with $v_{min}$ and so is maximized when
$v_{min}$=$v_{min}^{*Na}$ (corresponding for the benchmark point $P_1$
to $x_{min}^{*Na}\simeq$ 3.44 as shown in Fig.\ref{fig:xmin_eta1}). As
a consequence of this $\tilde{\eta}_1$ shows a maximum when
$E^{\prime}$=$E_{ee}^{*Na}$=$q E^{*Na}_R$ with $E^*_R$ defined in
Eq.(\ref{eq:estar}), i.e. the maximum in the modulation amplitude
detected by DAMA corresponds to WIMPs whose minimal incoming speed
matches the kinematic threshold of inelastic upscatters. In
particular, for the benchmark point $P_1$ one has $E_{ee}^*\simeq$
2.5, keVee and the fact that this value falls right in the energy
interval where the DAMA signal is detected is not a coincidence, but
descends from the requirement of Eq.(\ref{eq:hierarchy}). In this
case, however, the modulation amplitude at very low energies vanishes
instead of turning negative, because below $E^{*Na}_{ee}$ the halo
function $\tilde{\eta}_1(v_{min})$ is mapped in the same $v_{min}$
intervals than above $E^{*Na}_{ee}$, where
$\tilde{\eta}_1(v_{min})\ge$0: this may allow for a possible
discrimination among the two scenarios in a future low--threshold
analysis of the DAMA data \cite{dama_future}\footnote{The change of the
modulation amplitude can also be affected by the gravitational focusing
of the Sun\cite{focusing1,focusing2,focusing3}.}.

\section{Conclusions}
\label{sec:conclusions}

In the present paper we have discussed a scenario where the DAMA
modulation effect is explained by a WIMP which upscatters
inelastically to a heavier state and predominantly couples to the spin
of protons. In this scenario constraints from xenon and germanium
targets are evaded dynamically, due to the suppression of the WIMP
coupling $c^n$ to neutrons compared to the corresponding one $c^p$ to
protons and we have shown that, in order to evade the constraint from
LUX, the ratio $c^n/c^p$ must be tuned to a small but non--vanishing
value, which depends on which nuclear form factor is adopted. On the
other hand, the limits from fluorine targets are evaded by a
kinematical mechanism, because a combination of small values of the
WIMP mass $m_{DM}$ and large values of the mass splitting $\delta$
between the light and the heavy WIMP states drives the minimal WIMP
incoming speed $v_{min}^{*F}$ required to trigger upscatters off
fluorine either beyond the cut-off velocity $v_{cut}^{lab}$ of the
WIMP distribution $f(\vec{v_T})$ in the lab frame or very close to
it. On the other hand, WIMP scatterings off sodium can still explain
the yearly modulation effect in DAMA because the corresponding
$v_{min}^{*Na}$ is always smaller, since the sodium atomic mass is
larger than that of fluorine. In summary:
$v_{min}^{*Na}<v_{cut}^{lab}<v_{min}^{*F}$.  The corresponding
requirements for $m_{DM}$ and $\delta$ can be combined consistently to
those requiring that scatterings off iodine in COUPP are suppressed by
a similar kinematic mechanism ($v_{min}(E^{COUPP})>v_{cut}^{lab}$),
since the latter experiment can potentially constrain our scenario,
being sensitive to WIMP--proton scatterings and having a low recoil
energy threshold.

We have analyzed quantitatively the available data from direct
detection experiments either by fixing $f(\vec{v})$ to a Maxwellian
distribution and factorizing a reference WIMP--nucleon point--like
cross section $\sigma_{ref}$, or adopting instead a halo--independent
approach, where the dependence of the expected signal on the specific
choice of $f(\vec{v})$ is factorized in a halo function $\tilde{\eta}$
and its corresponding modulated one $\tilde{\eta}_1$, with the
requirement $\tilde{\eta}_1\le\tilde{\eta}$.  In the Maxwellian
analysis we have made the usual identification of $v_{cut}^{lab}$ with
the Galactic escape velocity boosted in the Earth's rest frame, while
in the halo--independent one we have instead made the conservative
assumption that $v_{cut}^{lab}$ is the maximal $v_{min}$ value
corresponding to the recoil energy range of the DAMA modulation
effect. The latter procedure corresponds to adopting the minimal halo
function explaining DAMA and compatible to the requirement of being a
decreasing function of $v_{min}$, and leads to laxer constraints in
the $m_{DM}$-$\delta$ plane when used to calculate expected signals in
null experiments.

The systematic scan of the $m_{DM}$--$\delta$ parameter space using a
quantitative compatibility factor ${\cal D}_{Maxwellian}$ for the
Maxwellian case and ${\cal D}$ for the halo--independent one (both
containing a full treatment of the expected rates in all experiments)
has shown that in both cases the kinematic requirements
$v_{min}^{*Na}<v_{cut}^{lab}<v_{min}^{*F}$,
$v_{min}(E^{COUPP})>v_{cut}^{lab}$ are sufficient to find allowed
configurations but not strictly necessary, since the allowed regions
are significantly larger than the kinematic ones, especially for the
halo--independent case. Our results are summarized in
Fig.\ref{fig:mchi_delta_full_analysis}, which shows that for the
Maxwellian case allowed configurations encompass the ranges 10
GeV$\lsim m_{DM}\lsim$ 13.5 GeV, 19.8 keV $\lsim \delta \lsim$ 26.5
keV, while in the case of a halo--independent analysis such region
enlarges considerably to 9.6 GeV $\lsim m_{DM}\lsim$ 31.3 GeV, 8.4 keV
$\lsim \delta \lsim$ 27.3 keV\footnote{In our analysis we have assumed
  that the DAMA signal is concentrated in the range 2 keVee$\le
  E_{ee}^{DAMA} \le 4$ keVee. Requiring a positive signal also in the
  energy range 4 keVee$\le E_{ee}^{DAMA} \le 6$ keVee would lower the
  kinematic upper--left boundaries of the allowed regions of
  Fig.\ref{fig:mchi_delta_full_analysis} without affecting the
  remaining parts.}.

We have shown that in our scenario the same kinematic mechanism that
suppresses the response to WIMP scatterings off fluorine compared to
sodium can lead in a natural way to a yearly modulation of the signal
in DAMA much larger than usually expected and which can approach 100\%
of the time--averaged part. This is always true for the allowed
configurations in the Maxwellian case, while configurations allowed by
the halo--independent treatment exist for which this effect may not be
present. However, at the boundary of the parameter region allowed by
the halo--independent analysis large modulation fractions are required
to evade the constraints on DAMA from other experiments.

Assuming a Maxwellian distribution we have shown that the large
modulated fractions of the signal arising in our scenario imply a
departure from the usual cosine time dependence, but even in the most
extreme cases the DAMA data is not sensitive to this distortion, both
in time and frequency space. On the other hand, when the modulated
fraction is large the signal contribution to the time-averaged DAMA
measured count--rates can be much smaller than usually assumed, easing
possible tension with DAMA background estimations existing in the
literature.

In our scenario the DAMA modulation effect can be explained in terms
of large values of $v_{min}$. In this regime the Maxwellian modulated
halo function $\tilde{\eta}_1$ decreases with $v_{min}$ and so is
maximized by $v_{min}$=$v^{*Na}_{min}$. Moreover, the requirement
$v_{min}^{*Na}<v_{cut}^{lab}<v_{min}^{*F}$ automatically implies that
the recoil energy $E_R^{*Na}$ which corresponds to $v_{min}^{*Na}$
falls in the DAMA energy range. This provides a simple explanation of
the maximum in the modulation amplitude detected by DAMA in terms of
WIMPs whose $v_{min}$ matches the kinematic threshold for inelastic
upscatters.  Interestingly, while for the elastic case the presence of
a maximum suggests an inversion of the modulation phase below the
present DAMA energy threshold, this does not happen for inelastic
scattering.  This may possibly allow for a discrimination between the
two scenarios in a future low--threshold analysis of the DAMA data.

\acknowledgments This work was supported by the National Research
Foundation of Korea(NRF) grant funded by the Korea government(MOE)
(No. 2011-0024836).

{\it Note added in proof.} After completing our analysis new improved
WIMP direct detection experimental constraints where published by LUX
\cite{lux_2016} and PICO-2L \cite{pico2l_2016}. As far as the new LUX
result is concerned it does not affect our conclusions due to the
Xe--phobic nature of our WIMP candidate for the choice of the
$c^n/c^p$ ratio adopted in our analysis. As far as the new PICO-2L
result is concerned, while the allowed kinematic regions described in
Section \ref{sec:fluorophobic} are necessarily not affected, since in
such kinematic ranges {\it any} fluorine target turns blind to WIMP
particles, its systematic inclusion in the extended numerical analysis
of Section \ref{sec:analysis} may somewhat affect the shape and
extension of the boundaries of the regions shown in
Fig.\ref{fig:mchi_delta_full_analysis}.

\appendix

\section{Experimental inputs}
\label{app:exp}

Compared to the analysis of Ref. \cite{eft_spin}, to which we refer
for all relevant details, in the present analysis we include two
additional experiments: the recent analyses by CDMS II
\cite{cdms_2015} and PICO-60 \cite{pico60}.

As far as the analysis of CDMS II is concerned, we consider only the
"classic" analysis with 10 keV threshold (see first entry of Table II
of Ref.\cite{cdms_2015}), with the two WIMP candidates shown in Table
I and the efficiency taken from the black solid line of the top panel
of Fig. 10 of the same paper.

PICO-60 is a bubble--chamber filled with 36.8 kg of $C F_3 I$. For
each of its operating thresholds (7 keV, 8.2 keV, 9.6 keV, 11.5 keV,
13.0 keV, 14.5 keV, 17.0 keV) we have used the exposures obtained by
combining the livetimes shown in Table 1 of Ref. \cite{pico60} with
the overall acceptance after cuts of 48.2\%.  No WIMP--event
candidates survive after the cuts, with a background estimation of 1.0
neutron--induced single bubble events and a 90\% C.L. upper bound of
2.33 events \cite{feldman_cousin}.


\begin{thebibliography}{99}
\bibitem{dama}
{\bf DAMA, LIBRA} Collaboration, R.~Bernabei et~al., {\it {New results from
  DAMA/LIBRA}},  {\em Eur. Phys. J.} {\bf C67} (2010) 39--49,
  [\href{http://arxiv.org/abs/1002.1028}{{\tt arXiv:1002.1028}}].

\bibitem{lux}
{\bf LUX} Collaboration, D.~S. Akerib et~al., {\it {First results from the LUX
  dark matter experiment at the Sanford Underground Research Facility}},  {\em
  Phys. Rev. Lett.} {\bf 112} (2014) 091303,
  [\href{http://arxiv.org/abs/1310.8214}{{\tt arXiv:1310.8214}}].

\bibitem{xenon100}
{\bf XENON100} Collaboration, E.~Aprile et~al., {\it {Dark Matter Results from
  225 Live Days of XENON100 Data}},  {\em Phys. Rev. Lett.} {\bf 109} (2012)
  181301, [\href{http://arxiv.org/abs/1207.5988}{{\tt arXiv:1207.5988}}].

\bibitem{xenon10}
{\bf XENON10} Collaboration, J.~Angle et~al., {\it {A search for light dark
  matter in XENON10 data}},  {\em Phys. Rev. Lett.} {\bf 107} (2011) 051301,
  [\href{http://arxiv.org/abs/1104.3088}{{\tt arXiv:1104.3088}}]. [Erratum:
  Phys. Rev. Lett.110,249901(2013)].

\bibitem{kims}
S.~C. Kim et~al., {\it {New Limits on Interactions between Weakly Interacting
  Massive Particles and Nucleons Obtained with CsI(Tl) Crystal Detectors}},
  {\em Phys. Rev. Lett.} {\bf 108} (2012) 181301,
  [\href{http://arxiv.org/abs/1204.2646}{{\tt arXiv:1204.2646}}].

\bibitem{kims_modulation} Y.~Kim, {\it {Recent progress in KIMS
    experiment}, talk given at {\it 13$^{th}$ International
    Conference on Topics in Astroparticle and Underground Physics,
    September 8--13 2013, Asilomar, California USA (TAUP2013)}}, .

\bibitem{kims2}
H.~S. Lee et~al., {\it {Search for Low-Mass Dark Matter with CsI(Tl) Crystal
  Detectors}},  {\em Phys. Rev.} {\bf D90} (2014), no.~5 052006,
  [\href{http://arxiv.org/abs/1404.3443}{{\tt arXiv:1404.3443}}].

\bibitem{cdms_ge}
{\bf CDMS-II} Collaboration, Z.~Ahmed et~al., {\it {Results from a Low-Energy
  Analysis of the CDMS II Germanium Data}},  {\em Phys. Rev. Lett.} {\bf 106}
  (2011) 131302, [\href{http://arxiv.org/abs/1011.2482}{{\tt
  arXiv:1011.2482}}].

\bibitem{cdms_lite}
{\bf SuperCDMS} Collaboration, R.~Agnese et~al., {\it {Search for Low-Mass
  Weakly Interacting Massive Particles Using Voltage-Assisted Calorimetric
  Ionization Detection in the SuperCDMS Experiment}},  {\em Phys. Rev. Lett.}
  {\bf 112} (2014), no.~4 041302, [\href{http://arxiv.org/abs/1309.3259}{{\tt
  arXiv:1309.3259}}].

\bibitem{super_cdms}
{\bf SuperCDMS} Collaboration, R.~Agnese et~al., {\it {Search for Low-Mass
  Weakly Interacting Massive Particles with SuperCDMS}},  {\em Phys. Rev.
  Lett.} {\bf 112} (2014), no.~24 241302,
  [\href{http://arxiv.org/abs/1402.7137}{{\tt arXiv:1402.7137}}].

\bibitem{cdms_2015}
{\bf SuperCDMS} Collaboration, R.~Agnese et~al., {\it {Improved WIMP-search
  reach of the CDMS II germanium data}},  {\em Phys. Rev.} {\bf D92} (2015),
  no.~7 072003, [\href{http://arxiv.org/abs/1504.05871}{{\tt
  arXiv:1504.05871}}].

\bibitem{simple}
M.~Felizardo et~al., {\it {Final Analysis and Results of the Phase II SIMPLE
  Dark Matter Search}},  {\em Phys. Rev. Lett.} {\bf 108} (2012) 201302,
  [\href{http://arxiv.org/abs/1106.3014}{{\tt arXiv:1106.3014}}].

\bibitem{coupp}
{\bf COUPP} Collaboration, E.~Behnke et~al., {\it {First Dark Matter Search
  Results from a 4-kg CF$_3$I Bubble Chamber Operated in a Deep Underground
  Site}},  {\em Phys. Rev.} {\bf D86} (2012), no.~5 052001,
  [\href{http://arxiv.org/abs/1204.3094}{{\tt arXiv:1204.3094}}]. [Erratum:
  Phys. Rev.D90,no.7,079902(2014)].

\bibitem{picasso}
{\bf PICASSO} Collaboration, S.~Archambault et~al., {\it {Constraints on
  Low-Mass WIMP Interactions on $^{19}F$ from PICASSO}},  {\em Phys. Lett.}
  {\bf B711} (2012) 153--161, [\href{http://arxiv.org/abs/1202.1240}{{\tt
  arXiv:1202.1240}}].

\bibitem{pico2l}
{\bf PICO} Collaboration, C.~Amole et~al., {\it {Dark Matter Search Results
  from the PICO-2L C$_3$F$_8$ Bubble Chamber}},  {\em Phys. Rev. Lett.} {\bf
  114} (2015), no.~23 231302, [\href{http://arxiv.org/abs/1503.00008}{{\tt
  arXiv:1503.00008}}].

\bibitem{pico60}
{\bf PICO} Collaboration, C.~Amole et~al., {\it {Dark Matter Search Results
  from the PICO-60 CF$_3$I Bubble Chamber}},  {\em Submitted to: Phys. Rev. D}
  (2015) [\href{http://arxiv.org/abs/1510.07754}{{\tt arXiv:1510.07754}}].

\bibitem{spin_n_suppression}
P.~Ullio, M.~Kamionkowski, and P.~Vogel, {\it {Spin dependent WIMPs in DAMA?}},
   {\em JHEP} {\bf 07} (2001) 044,
  [\href{http://arxiv.org/abs/hep-ph/0010036}{{\tt hep-ph/0010036}}].

\bibitem{spin_gelmini}
E.~Del~Nobile, G.~B. Gelmini, A.~Georgescu, and J.-H. Huh, {\it {Reevaluation
  of spin-dependent WIMP-proton interactions as an explanation of the DAMA
  data}},  {\em JCAP} {\bf 1508} (2015), no.~08 046,
  [\href{http://arxiv.org/abs/1502.07682}{{\tt arXiv:1502.07682}}].

\bibitem{spin_freitsis}
M.~Freytsis and Z.~Ligeti, {\it {On dark matter models with uniquely
  spin-dependent detection possibilities}},  {\em Phys. Rev.} {\bf D83} (2011)
  115009, [\href{http://arxiv.org/abs/1012.5317}{{\tt arXiv:1012.5317}}].

\bibitem{spin_arina}
C.~Arina, E.~Del~Nobile, and P.~Panci, {\it {Dark Matter with
  Pseudoscalar-Mediated Interactions Explains the DAMA Signal and the Galactic
  Center Excess}},  {\em Phys. Rev. Lett.} {\bf 114} (2015) 011301,
  [\href{http://arxiv.org/abs/1406.5542}{{\tt arXiv:1406.5542}}].

\bibitem{eft_spin}
S.~Scopel, K.-H. Yoon, and J.-H. Yoon, {\it {Generalized spin-dependent
  WIMP-nucleus interactions and the DAMA modulation effect}},  {\em JCAP} {\bf
  1507} (2015), no.~07 041, [\href{http://arxiv.org/abs/1505.01926}{{\tt
  arXiv:1505.01926}}].

\bibitem{inelastic}
D.~Tucker-Smith and N.~Weiner, {\it {Inelastic dark matter}},  {\em Phys. Rev.}
  {\bf D64} (2001) 043502, [\href{http://arxiv.org/abs/hep-ph/0101138}{{\tt
  hep-ph/0101138}}].

\bibitem{v_loc}
 We adopt here the conventional value $v_{loc}$=220 km/s which is used in most
  Dark Matter direct detection experimental papers. For a more recent
  determination of $v_{loc}$ see, for instance, Ref. \cite{vesc_salucci}.

\bibitem{inverse_modulation}
M.~J. Lewis and K.~Freese, {\it {The Phase of the annual modulation:
  Constraining the WIMP mass}},  {\em Phys. Rev.} {\bf D70} (2004) 043501,
  [\href{http://arxiv.org/abs/astro-ph/0307190}{{\tt astro-ph/0307190}}].

\bibitem{dama_future}
R.~Bernabei et~al., {\it {New Results from DAMA/LIBRA: Final Model-Independent
  Results of Dama/Libra-Phase1 and Perspectives of Phase2}},  {\em Frascati
  Phys. Ser.} {\bf 58} (2014) 41.

\bibitem{factorization}
P.~J. Fox, J.~Liu, and N.~Weiner, {\it {Integrating Out Astrophysical
  Uncertainties}},  {\em Phys. Rev.} {\bf D83} (2011) 103514,
  [\href{http://arxiv.org/abs/1011.1915}{{\tt arXiv:1011.1915}}].

\bibitem{mccabe_eta}
C.~McCabe, {\it {DAMA and CoGeNT without astrophysical uncertainties}},  {\em
  Phys. Rev.} {\bf D84} (2011) 043525,
  [\href{http://arxiv.org/abs/1107.0741}{{\tt arXiv:1107.0741}}].

\bibitem{gondolo_eta1}
P.~Gondolo and G.~B. Gelmini, {\it {Halo independent comparison of direct dark
  matter detection data}},  {\em JCAP} {\bf 1212} (2012) 015,
  [\href{http://arxiv.org/abs/1202.6359}{{\tt arXiv:1202.6359}}].

\bibitem{gondolo_eta2}
E.~Del~Nobile, G.~B. Gelmini, P.~Gondolo, and J.-H. Huh, {\it {Halo-independent
  analysis of direct detection data for light WIMPs}},  {\em JCAP} {\bf 1310}
  (2013) 026, [\href{http://arxiv.org/abs/1304.6183}{{\tt arXiv:1304.6183}}].

\bibitem{gondolo_eta3}
E.~Del~Nobile, G.~B. Gelmini, P.~Gondolo, and J.-H. Huh, {\it {Update on Light
  WIMP Limits: LUX, lite and Light}},  {\em JCAP} {\bf 1403} (2014) 014,
  [\href{http://arxiv.org/abs/1311.4247}{{\tt arXiv:1311.4247}}].

\bibitem{noi1}
S.~Scopel and K.~Yoon, {\it {A systematic halo-independent analysis of direct
  detection data within the framework of Inelastic Dark Matter}},  {\em JCAP}
  {\bf 1408} (2014) 060, [\href{http://arxiv.org/abs/1405.0364}{{\tt
  arXiv:1405.0364}}].

\bibitem{freese}
K.~Freese, M.~Lisanti, and C.~Savage, {\it {Colloquium: Annual modulation of
  dark matter}},  {\em Rev. Mod. Phys.} {\bf 85} (2013) 1561--1581,
  [\href{http://arxiv.org/abs/1209.3339}{{\tt arXiv:1209.3339}}].

\bibitem{vesc}
M.~C. Smith et~al., {\it {The RAVE Survey: Constraining the Local Galactic
  Escape Speed}},  {\em Mon. Not. Roy. Astron. Soc.} {\bf 379} (2007) 755--772,
  [\href{http://arxiv.org/abs/astro-ph/0611671}{{\tt astro-ph/0611671}}].

\bibitem{vesc_salucci}
F.~Nesti and P.~Salucci, {\it {The Dark Matter halo of the Milky Way, AD
  2013}},  {\em JCAP} {\bf 1307} (2013) 016,
  [\href{http://arxiv.org/abs/1304.5127}{{\tt arXiv:1304.5127}}].

\bibitem{self_truncated}
S.~Chaudhury, P.~Bhattacharjee, and R.~Cowsik, {\it {Direct detection of WIMPs
  : Implications of a self-consistent truncated isothermal model of the Milky
  Way's dark matter halo}},  {\em JCAP} {\bf 1009} (2010) 020,
  [\href{http://arxiv.org/abs/1006.5588}{{\tt arXiv:1006.5588}}].

\bibitem{haxton1}
A.~L. Fitzpatrick, W.~Haxton, E.~Katz, N.~Lubbers, and Y.~Xu, {\it {The
  Effective Field Theory of Dark Matter Direct Detection}},  {\em JCAP} {\bf
  1302} (2013) 004, [\href{http://arxiv.org/abs/1203.3542}{{\tt
  arXiv:1203.3542}}].

\bibitem{haxton2}
N.~Anand, A.~L. Fitzpatrick, and W.~C. Haxton, {\it {Weakly interacting massive
  particle-nucleus elastic scattering response}},  {\em Phys. Rev.} {\bf C89}
  (2014), no.~6 065501, [\href{http://arxiv.org/abs/1308.6288}{{\tt
  arXiv:1308.6288}}].

\bibitem{noi2}
S.~Scopel and J.-H. Yoon, {\it {Effective scalar four-fermion interaction for
  Ge-phobic exothermic dark matter and the CDMS-II Silicon excess}},  {\em
  Phys. Rev.} {\bf D91} (2015), no.~1 015019,
  [\href{http://arxiv.org/abs/1411.3683}{{\tt arXiv:1411.3683}}].

\bibitem{halo_independent_inelastic}
N.~Bozorgnia, J.~Herrero-Garcia, T.~Schwetz, and J.~Zupan, {\it
  {Halo-independent methods for inelastic dark matter scattering}},  {\em JCAP}
  {\bf 1307} (2013) 049, [\href{http://arxiv.org/abs/1305.3575}{{\tt
  arXiv:1305.3575}}].

\bibitem{spin_form_factors}
M.~T. Ressell and D.~J. Dean, {\it {Spin dependent neutralino - nucleus
  scattering for A approximately 127 nuclei}},  {\em Phys. Rev.} {\bf C56}
  (1997) 535--546, [\href{http://arxiv.org/abs/hep-ph/9702290}{{\tt
  hep-ph/9702290}}].

\bibitem{non_thermal}
M.~Vogelsberger, A.~Helmi, V.~Springel, S.~D.~M. White, J.~Wang, C.~S. Frenk,
  A.~Jenkins, A.~D. Ludlow, and J.~F. Navarro, {\it {Phase-space structure in
  the local dark matter distribution and its signature in direct detection
  experiments}},  {\em Mon. Not. Roy. Astron. Soc.} {\bf 395} (2009) 797--811,
  [\href{http://arxiv.org/abs/0812.0362}{{\tt arXiv:0812.0362}}].

\bibitem{dama_bck}
V.~A. Kudryavtsev, M.~Robinson, and N.~J.~C. Spooner, {\it {The expected
  background spectrum in NaI dark matter detectors and the DAMA result}},  {\em
  Astropart. Phys.} {\bf 33} (2010) 91--96,
  [\href{http://arxiv.org/abs/0912.2983}{{\tt arXiv:0912.2983}}].

\bibitem{focusing1}
S.~K. Lee, M.~Lisanti, A.~H.~G. Peter, and B.~R. Safdi, {\it {Effect of
  Gravitational Focusing on Annual Modulation in Dark-Matter Direct-Detection
  Experiments}},  {\em Phys. Rev. Lett.} {\bf 112} (2014), no.~1 011301,
  [\href{http://arxiv.org/abs/1308.1953}{{\tt arXiv:1308.1953}}].

\bibitem{focusing2}
N.~Bozorgnia and T.~Schwetz, {\it {Is the effect of the Sun's gravitational
  potential on dark matter particles observable?}},  {\em JCAP} {\bf 1408}
  (2014) 013, [\href{http://arxiv.org/abs/1405.2340}{{\tt arXiv:1405.2340}}].

\bibitem{focusing3}
E.~Del~Nobile, G.~B. Gelmini, and S.~J. Witte, {\it {Gravitational Focusing and
  Substructure Effects on the Rate Modulation in Direct Dark Matter Searches}},
   {\em JCAP} {\bf 1508} (2015), no.~08 041,
  [\href{http://arxiv.org/abs/1505.07538}{{\tt arXiv:1505.07538}}].

\bibitem{lux_2016}
{\bf LUX} Collaboration, D.~S. Akerib et~al., {\it {Improved WIMP scattering
  limits from the LUX experiment}},
  \href{http://arxiv.org/abs/1512.03506}{{\tt arXiv:1512.03506}}.

\bibitem{pico2l_2016}
{\bf PICO} Collaboration, C.~Amole et~al., {\it {Improved Dark Matter Search
  Results from PICO-2L Run-2}},  \href{http://arxiv.org/abs/1601.03729}{{\tt
  arXiv:1601.03729}}.

\bibitem{feldman_cousin}
G.~J. Feldman and R.~D. Cousins, {\it {A Unified approach to the classical
  statistical analysis of small signals}},  {\em Phys. Rev.} {\bf D57} (1998)
  3873--3889, [\href{http://arxiv.org/abs/physics/9711021}{{\tt
  physics/9711021}}].

\end{thebibliography}
\end{document}